\documentstyle[12pt,emlines]{article}
\def\appendix#1{
\addtocounter{section}{1}
\setcounter{equation}{0}
\renewcommand{\thesection}{\Alph{section}}
\section*{Appendix \thesection\protect\indent #1}
\addcontentsline{toc}{section}{Appendix \thesection\ \ \ #1}
}

\def\e{{\,\rm e}\,}

\def\be{\begin{equation}}
\def\la{\label}
\def\ee{\end{equation}}
\def\bea{\begin{eqnarray}}
\def\eea{\end{eqnarray}}

\def\a{\alpha}

\def\s{\sigma}
\def\D{\Delta}
\def\d{\delta}
\newcommand{\R}{{\bf R}}
\newcommand{\k}{{\bf k}}
\newcommand{\q}{{\bf q}}
\newcommand{\Z}{{\bf Z}}
\newcommand{\cH}{{\cal H }}
 \newcommand{\cL}{{\cal L }}
\begin{document}
\title{\hfill{LMU-TPW 97-21} \\
\vspace{1cm}
Virasoro amplitude from the $S^N\large{{\bf R}}^{24}$ orbifold sigma model}
\author{G.E.Arutyunov
\thanks{arut@genesis.mi.ras.ru}
\mbox{} \\
\vspace{0.4cm}
Steklov Mathematical Institute,
\vspace{-0.5cm} \mbox{} \\
Gubkin str.8, GSP-1, 117966, Moscow, Russia;
\vspace{0.5cm} \mbox{} \\
S.A.Frolov\thanks{Alexander von Humboldt fellow}
\mbox{} \\
\vspace{0.4cm}
Section Physik, Munich University
\vspace{-0.5cm} \mbox{} \\
Theresienstr.37, 80333 Munich, Germany
\thanks{Permanent address:\ Steklov Mathematical Institute, Moscow}
\mbox{}}
\date {} \maketitle \begin{abstract}
Four tachyon scattering amplitude is derived from the $S^N\R^{24}$ orbifold sigma
model in the large $N$ limit. The closed string interaction is described by a
vertex which is a bosonic analog of the supersymmetric one, recently proposed
by Dijkgraaf, Verlinde and Verlinde.
\end{abstract}

\section{Introduction}
Compactification of M(atrix) theory \cite{BFSS} on a circle results in the
${\cal N}=8$ two-dimensional supersymmetric $SU(N)$ Yang-Mills model \cite{T}.
It was recently suggested in \cite{M,BS,DVV} that in the large $N$ limit the
Yang-Mills theory describes non-perturbative dynamics of type IIA string
theory, and the string coupling constant was argued to be inverse proportional
to the Yang-Mills coupling. This suggestion looks very natural since
in the IR limit the gauge theory is strongly coupled and the IR fixed point
may be described by the ${\cal N}=8$ supersymmetric conformal field theory
on the orbifold target space $S^N\R^8$.  The Hilbert space of
the orbifold model is known \cite{DMVV} to coincide (to be precise, to contain)
in the large $N$ limit with the Fock space of the free second-quantized type
IIA string theory. Using these facts, Dijkgraaf, Verlinde and Verlinde (DVV)
\cite{DVV} have suggested that perturbative string dynamics in the first order
in the string coupling constant can be described by the $S^N\R^{8}$
supersymmetric orbifold conformal model perturbed by an irrelevant operator of
conformal dimension $(3/2,3/2)$. An explicit form of this operator $V$ was
determined in \cite{DVV} and it nicely fits the conventional formalism of the
light-cone string theory.

The described approach  seems not to be limited only to the
supersymmetric case. In particular, one can suggest \cite{R} that the
M(atrix) theory formulation for closed bosonic strings is provided
by the large $N$ limit of the two-dimensional Yang-Mills theory with
$24$ matter fields in the adjoint representation of the $U(N)$ gauge group.
In this case, the IR limit of the gauge theory results in
the $S^{N}\R^{24}$ orbifold conformal model. The closed bosonic
string interactions are described via perturbation of the CFT action
with a bosonic analog of the DVV vertex \cite{R}.

An important problem posed by the above-described stringy
interpretation of the $S^N$ orbifold sigma models is to obtain the
usual string scattering amplitudes directly from the models. This problem seems
to be nontrivial due to the nonabelian nature of the $S^N$ orbifold models.

The aim of the present paper is to derive the four tachyon scattering
amplitude from the $S^{N}\R^{24}$ orbifold conformal field theory
perturbed by the bosonic analog of the DVV interaction vertex.

Obviously, the first step in constructing the scattering amplitudes
consists in defining the incoming and outgoing asymptotic states
$|i>$ and $|f>$. The free string limit $g_s\to 0$
implies that the asymptotic states should be identified with
some states in the Hilbert space of the orbifold conformal field
theory and, therefore, they should be created by some conformal
fields. Then, by the conventional quantum field theory, the
$g_s^n$-order scattering amplitude $A$ can be extracted from the S-matrix
element described as a correlator of $n$ conformal fields $V(z_i)$ with the
subsequent integration over the insertion points $z_i$:
$$
<f|S|i>\sim \int
\prod_{i}d^2z_i <f|V(z_1)\ldots V(z_n)|i>.
$$
The construction of the asymptotic states $|i>$ and $|f>$
that can be identified with incoming and outgoing tachyons,
and computation of the above-mentioned correlators in the
$S^{N}\R^{24}$ orbifold CFT are the main questions we are dealing with in the
paper to obtain the four tachyon scattering amplitude.

The paper is organized as follows. In the second section we remind the
description of the Hilbert space of the orbifold model. In the third section
the twist fields that create the states of the Hilbert space are introduced
and their conformal dimensions are calculated. In the fourth section the
scattering amplitude is calculated and is shown to coincide with the Virasoro
one. In Conclusion we discuss unsolved problems.

\section{$S^N\R^{D}$ orbifold sigma model}
We consider two-dimensional field theory on a cylinder described by the action
\be
S=\frac{1}{2\pi}\int d\tau d\sigma  (\partial_{\tau}X^i_I\partial_{\tau}X^i_I -
\partial_{\sigma}X^i_I\partial_{\sigma}X^i_I ),
\label{act}
\ee
where $0\leq \sigma < 2\pi$, $i=1,2,\dots ,D$, $I=1,2,\dots ,N$ and the
fields $X$ take values in $S^N\R^{D}\equiv (\R^{D})^N/S_N$.

As usual in orbifold models \cite{DHVW1,DHVW2}, the field $X^i$ can have twisted
boundary conditions
\be
X^i(\sigma +2\pi )=gX^i(\sigma ),
\label{bc}
\ee
where $g$ belongs to the symmetric group $S_N$.

\noindent Multiplying (\ref{bc}) by some element $h\in S_N$ and taking into
account that $X^i$ and $hX^i$ describe the same configuration, one gets that all
possible boundary conditions are in one-to-one correspondence with the
conjugacy classes of the symmetric group. Therefore, the Hilbert space of the
orbifold model is decomposed into the direct sum of Hilbert spaces of the
twisted sectors corresponding to the conjugacy classes $[g]$ of $S_N$
\cite{DMVV}
\bea
\cH(S^N\R^{D}) = \bigoplus_{[g]} \cH_{[g]}.
\nonumber
\eea
It is well-known that the conjugacy classes of $S_N$ are described by
partitions $\{ N_n\}$ of $N$
\bea
N=\sum_{n=1}^s\, nN_n
\nonumber
\eea
and can be represented as
\be
[g] = (1)^{N_1}(2)^{N_2} \cdots (s)^{N_s}.
\label{facg}
\ee
Here $N_n$ is the multiplicity of the cyclic permutation $(n)$
of $n$ elements.

In any conjugacy class $[g]$ there is the only element $g_c$ that has  the
canonical block-diagonal form
\be
g_c=diag(\underbrace{\omega _1,...,\omega _1}_{N_1\ times},
\underbrace{\omega _2,...,\omega _2}_{N_2\ times},...,
\underbrace{\omega _s,...,\omega _s}_{N_s\ times}),
\label{canform}
\ee
where $\omega _n$ is an $n\times n$ matrix that generates the cyclic
permutation $(n)$ of $n$ elements
\bea
\omega _n=\sum _{i=1}^{n-1}E_{i,i+1} +E_{n1}
\nonumber
\eea
and $E_{ij}$ are matrix unities.

\noindent It is not difficult to show that $\omega _n$ generates the $\Z_n$
group, since $\omega _n^n=1$, and that only the matrices $\omega _n^k$ from
$\Z_n$ commute with $\omega _n$.
Since the centralizer subgroup $C_g$ of any element $g\in [g]$ is isomorphic
to $C_{g_c}$ one concludes that
\bea
C_g =
\prod_{n=1}^s \, S_{N_n} \times \Z^{N_n}_n,
\nonumber
\eea
where the symmetric group $S_{N_n}$ permutes the $N_n$ cycles $(n)$. It is
obvious that the stabilizer $C_g$ contains $\prod_{n=1}^s \, N_n!n^{N_n}$
elements.

Due to the factorization (\ref{facg}) of $[g]$, the Hilbert space
$\cH_{[g]}\equiv\cH_{\{ N_n\}}$ of each twisted sector can be decomposed into
the $N_n$-fold symmetric tensor products of the Hilbert spaces $\cH_{(n)}$
which correspond to the cycles of length $n$
\bea
{\cH}_{\{N_n\}} = \bigotimes_{n=1}^s \, S^{N_n} \cH_{(n)}
=\bigotimes_{n=1}^s \,\left(\underbrace{\cH_{(n)} \otimes \cdots
\otimes \cH_{(n)}}_{N_n\ times}\right)^{S_{N_n}}.
\nonumber\eea
The space $\cH_{(n)}$ is $\Z_n$ invariant subspace of the Hilbert space of a
sigma model of $Dn$ fields $X_I^i$ with the cyclic boundary condition
\be
X_I^i(\s +2\pi )=X_{I+1}^i(\s ),\quad I=1,2,...,n.
\label{cyc}
\ee
The fields $X_I(\s )$ can be glued together into one field $X(\s )$ that is
identified with a long string of the length $n$. The states of the space
$\cH_{(n)}$ are obtained by acting by the creation operators of the string on
eigenvectors of the momentum operator. These eigenvectors have the standard
normalization
\bea
\langle \q |\k\rangle =\d ^D(\q +\k )
\nonumber\eea
and can be regarded as states obtained by acting by the operator $\e ^{i\k x}$ on
the vacuum state (that is not normalizable):
$|\k\rangle =\e ^{i\k x}|0\rangle$, $\langle \q |=\langle 0 |\e ^{i\q x}$.

The $\Z_n$ invariant subspace  is
singled out by imposing the condition
\bea
(L_0-\bar {L}_0)|\Psi\rangle =nm|\Psi\rangle ,
\nonumber\eea
where $m$ is an integer and $L_0$ is the canonically normalized $L_0$ operator
of the single string.

If $D=24$ then the Fock space of the second-quantized closed bosonic string is
recovered in
the limit $N\to\infty$, $\frac {n_i}{N}\to p^+_i$ \cite{DMVV}, where the finite
ratio  $\frac {n_i}{N}$ is identified with the $p^+_i$ momentum of a long
string. The $\Z_n$ projection reduces in this limit to the usual level-matching
condition $L_0^{(i)}-\bar {L}_0^{(i)}=0$. The individual $p^-_i$ light-cone
momentum is defined by means of the standard mass-shell condition
$p^+_ip^-_i=L_0^{(i)}$.

\section{Twist fields}
Let us consider conformal field theory of $DN$ free scalar fields described by
the action (\ref{act}). It is convinient to perform the Wick rotation
$\tau\to -i\tau$ and to map the cylinder onto the sphere:
$z=\e ^{\tau +i\s }$, $\bar {z}=\e ^{\tau -i\s }$.

The vacuum state $|0\rangle$ of the CFT is annihilated by the momentum operators and by
annihilation operators, and has to be normalizable. To be able to identify this vacuum
state with the vacuum state of the untwisted sector of the orbifold sigma model
we choose the following normalization of $|0\rangle$
\bea
\langle 0 |0\rangle =R^{DN}.
\nonumber\eea
Here $R$ should be regarded as a regularization parameter of the sigma model.
We regularize the sigma model by compactifying the coordinates $x^i_I$ on
circles of radius $R$. Then the norm of the eigenvectors of the momentum
operators in the untwisted sector is given by
\bea
\langle \q |\k\rangle =(2\pi )^{-DN}\int_0^{2\pi R}d^{DN}x \e ^{i(\q +\k )x}=
\prod_{I=1}^N \d ^D_R(\q_I +\k_I ),
\nonumber\eea
where $k^i_I=\frac {m^i_I}{R}$ and $q^i_I=\frac {n^i_I}{R}$ are momenta of the
states, $m^i_I$ and $n^i_I$ are integers since we compactified the coordinates,
and $\d ^D_R(\k )=R^D\prod_{i=1}^D\d _{m^i0}$ is the regularized $\d$-function.
In the limit $R\to\infty$ one recovers the usual normalization of the
eigenvectors.

As usual, the field $X(z,\bar {z})$ can be decomposed into the left- and
right-moving components
\be
2X(z,\bar {z})=X(z)+\bar {X}(\bar {z}).
\label{xdec}
\ee
In what follows we shall mainly concentrate our attention on the left-moving
sector.

\noindent Let $\s _g(z,\bar {z})$ be a primary field \cite{BPZ} that creates a
vacuum of a twisted sector at the point $z$, i.e. the fields
$X^i(z,\bar {z})$ satisfy the following monodromy conditions
\bea
X^i(z\e^{2\pi i},\bar {z}\e^{-2\pi i})\s _g(0,0) =gX^i(z,\bar {z})\s _g(0,0).
\nonumber\eea
It is clear that the twist field $\s _g(z,\bar {z})$ can be represented as the
tensor product of the twist fields $\s _g(z)$ and $\bar \s _g(\bar {z})$ that
create the vacuum states of the left- and right-moving sectors respectively:
$\s _g(z,\bar {z})=\s _g(z)\otimes\bar \s _g(\bar {z})$.

It is obvious that the conformal dimension $\D _g$ depends only on $[g]$. To
calculate $\D _g$ let us suppose that $g$ has the factorization (\ref{facg}).
Then $\D _g$ is given by the equation
\be
\D _g=\sum_{n=1}^s\, N_n\D _{(n)},
\label{confdim}
\ee
where $\D _{(n)}$ denotes the conformal dimension of the twist field
$\s _{(n)}$ that creates the vacuum state of the space $\cH_{(n)}$ of the sigma
model of $Dn$ fields with cyclic boundary condition (\ref{cyc}). Let the twist
field $\s _{(n)}$ be located at $z=0$ and let us denote the vacuum
state \footnote{This vacuum state is a primary state of the CFT.} as
$|(n)\rangle =\s _{(n)}(0)|0\rangle$. Since the twist field
$\s _{(n)}$ creates one long string we normalize the vacuum state $|(n)\rangle$
as
\be
\langle (n) |(n)\rangle =R^{D}.
\label{normvac1}
\ee

The fields $X(z)$ have the following
decomposition in the vicinity of $z=0$
\be
\partial X_I^i(z)=-i\frac 1 n \sum_m \a _m^i\e^{-\frac {2\pi i}{n}Im}
z^{-\frac {m}{n}-1},
\label{moddec}
\ee
where $\a _m^i$ ($m\neq 0$) are the usual creation and annihilation operators
with the commutation relations
\be
[\a _m^i, \a _n^j]=m\d ^{ij}\d _{m+n,0},
\label{comrel}
\ee
and $\a _0^i$ is proportional to the momentum operator
\footnote{$\a _0^i=\frac{1}{2}p^i$ in string units $\alpha'=\frac{1}{2}$.}.

\noindent The vacuum state $|(n)\rangle $ is annihilated by the operators
$\a _m^i$ for $m\geq 0$.

Since $\s _{(n)}$ is a primary field, the conformal dimension $\D _{(n)}$ can be
found from the equation
\bea
\langle (n)|T(z)|(n)\rangle =\frac {\D _{(n)}}{z^2}\langle (n)|(n)\rangle ,
\nonumber
\eea
where $T(z)$ is the stress-energy tensor.

\noindent By using eqs. (\ref{moddec}) and (\ref{comrel}), one calculates the
correlator
\bea
\langle (n)|\partial X_I^i(z)\partial X_I^j(w)|(n)\rangle
=-\d ^{ij}\frac {(zw)^{\frac 1 n -1}}{n^2(z^{\frac 1 n} -w^{\frac 1 n})^2}
\langle (n)|(n)\rangle .
\nonumber
\eea
Taking into account that the stress-energy tensor is defined as
\bea
T(z)= -\frac 12 \lim _{w\to z}\sum_{i=1}^{D}\sum_{I=1}^n\left(
\partial X_I^i(z)\partial X_I^i(w) +\frac {1}{(z-w)^2}\right) ,
\nonumber
\eea
one gets
\be
\D _{(n)}=\frac {D}{24} (n-\frac {1} {n} ).
\label{confdim1}
\ee
The excited states of this sigma model are obtained by acting on $|(n)\rangle $
by some vertex operators. In particular the state corresponding to a scalar
particle with momentum $\k$ is given by
\be
\s _{(n)}[\k](0,0)|0\rangle =:\e ^{ik^i_IX_I^i(0,0)}:|(n)\rangle ,
\label{tachst}
\ee
where the summation over $i$ and $I$ is assumed, $k^i_I=\frac {m^i_I}{R}$ is a momentum carried
by the field $X_I^i(z,\bar z)$ and $k^i=\sum_{I=1}^n k^i_I$ is a total momentum
of the long string.

\noindent By using the definition of the vacuum state $|(n)\rangle$, one can
rewrite eq.(\ref{tachst}) in the form
\be
\s _{(n)}[\k ](0,0)|0\rangle =:\e ^{i\frac {k^i}{\sqrt {n}}Y^i(0,0)}:
|(n)\rangle ,
\label{tachst1}
\ee
where
\be
Y^i(z,\bar z)=\frac {1}{\sqrt {n}}\sum_{I=1}^n X_I^i(z,\bar z).
\label{defY}
\ee
The field $Y(z)$ is canonically normalized, i.e. the part of the stress-energy
tensor depending on $Y$ is $-\frac 12 :\partial Y(z)\partial Y(z):$, and has
the trivial monodromy around $z=0$.

It is obvious from eq.(\ref{tachst1}) that the conformal dimension of the
primary field
$$
\s _{(n)}[\k ](z,\bar z) =
:\e ^{i\frac {k^i}{\sqrt {n}}Y^i(z,\bar z)}:\s _{(n)}(z,\bar z)
$$
is equal to
\bea
\D _{(n)}[\k ]=\D _{(n)}+\frac {\k ^2}{8n}=
\frac {D}{24} (n-\frac 1n)+\frac {\k ^2}{8n},
\nonumber
\eea
where the decomposition (\ref{xdec}) was taken into account.

Due to eqs. (\ref{confdim}) and (\ref{confdim1}), the  conformal dimension of
$\s _g$ is given by
\bea
\D _g=\sum_{n=1}^s\, N_n\frac {D}{24} (n-\frac 1n)=
\frac {D}{24}(N-\sum_{n=1}^s\, \frac {N_n}{n}).
\nonumber\eea
One can also introduce a primary field that creates scalar particles with
momenta $k^i_\a$, $\a =1,2,...,N_1+N_2+\cdots +N_s\equiv N_{str}$
\bea
\s _{g}[\{\k _\a\}](z,\bar z) =
:\e ^{i\frac {k^i_\a}{\sqrt {n_\a}}Y^i_\a (z,\bar z)}:\s _{g}(z,\bar z),
\nonumber\eea
where $n_1=n_2=\cdots =n_{N_1}=1$, $n_{N_1+1}=n_{N_1+2}=\cdots =n_{N_1+N_2}=2$
and so on, $Y^i_\a$ corresponds to the cycle $(n_\a )$ and is defined by
eq.(\ref{defY}), and the summation over $i$ and $\a$ is assumed.

\noindent The conformal dimension of the field $\s _{g}[\{\k _\a\}]$ is equal
to
\bea
\D _g[\{\k _\a\}]=
\frac {D}{24}(N-\sum_{n=1}^s\, \frac {N_n}{n})+\sum_\a
\frac {\k^2_\a}{8n_\a}.
\eea
It is obvious that the two-point correlation function of the twist fields
$\s_{g_1}$ and $\s_{g_2}$ is not equal to zero if and only if $g_1g_2=1$.
Taking into account the
normalization (\ref{normvac1}), we find
\footnote{it is clear that $[g^{-1}]=[g]$ and therefore $\D _{g^{-1}}=\D _{g}$}
\bea
\langle \s_{g^{-1}}(\infty)\s_g(0)\rangle =R^{DN_{str}}.
\nonumber
\eea
It means that the fields $\s_{g^{-1}}$ and $\s_g$ have the following OPE
\bea
\s_{g^{-1}}(z,\bar z)\s_g(0,0)=
\frac {R^{D(N_{str}-N)}}{|z|^{4\D _{g}}}+\cdots .
\nonumber
\eea
The two-point correlation function of $\s _{g^{-1}}[\{\q _\a\}]$ and
$\s _{g}[\{\k _\a\}]$ is respectively equal to
\be
\langle \s_{g^{-1}}[\{\q _\a\}](\infty)\s_g[\{\k _\a\}](0)\rangle =
\prod_\a  \d ^D_R(\q_\a +\k_\a ).
\label{norm}
\ee

The twist fields $\s _g$ do not create twisted sectors of the orbifold CFT
since they are not invariant with respect to the action of the symmetric group.
An invariant twist field can be defined by summing up all the twist fields from
one conjugacy class
\bea
\s _{[g]}(z,\bar z)=\frac {1}{N!}\sum_{h\in S_N}\s_{h^{-1}gh}(z,\bar z).
\nonumber\eea
By using this definition, one can easily calculate the two-point correlation
function
\bea
\langle \s_{[g]}(\infty)\s_{[g]}(0)\rangle =\frac {R^{DN_{str}}}{N!}
\prod_{n=1}^sN_n!n^{N_n},
\nonumber\eea
where $\prod_{n=1}^sN_n!n^{N_n}$ is the number of elements of the centralizer
subgroup $C_g$.

The definition of the twist field $\s _{[g]}[\{\k _\a\}]$ is not so
straightforward. Let us consider the element $g_c\in [g]$ that has the
canonical block-diagonal form (\ref{canform}).
There are $N_1+\cdots +N_s=N_{str}$ fields $Y_\a (z,\bar z)$ that have the
trivial monodromy in the vicinity of the location of the twist field
$\s _{g_c}$.
According to eq.(\ref{defY}) they are defined as
\bea
Y_\a (z,\bar z)=\frac {1}{\sqrt {n_\a }}\sum_{I\in (n_\a )}X_I(z,\bar z).
\nonumber\eea
Let us now consider the fields $X$ which have the monodromy
\be
X(z\e^{2\pi i},\bar {z}\e^{-2\pi i})=h^{-1}g_chX(z,\bar {z}).
\label{mon1}
\ee
One can see from eq.(\ref{mon1}) that the fields $Y_\a [h]$
\bea
Y_\a [h](z,\bar z)=\frac {1}{\sqrt {n_\a }}\sum_{I\in (n_\a )}(hX)_I(z,\bar z)
\nonumber
\eea
have the trivial monodromy.

\noindent Then an invariant twist field $\s _{[g]}[\{\k _\a\}]$ is defined as
follows
\be
\s _{[g]}[\{\k _\a\}](z,\bar z) =\frac {1}{N!}\sum_{h\in S_N}
:\e ^{i\frac {k^i_\a}{\sqrt {n_\a}}Y^i_\a [h](z,\bar z)}:
\s _{h^{-1}g_ch}(z,\bar z).
\label{tachop1}
\ee
One can easily check that the twist field $\s _{[g]}[\{\k _\a\}]$ is invariant
with respect to the permutation of momenta $\k _\a$ which correspond to cycles
$(n_\a )$ of the same length.

The interaction vertex proposed by DVV \cite{DVV} is defined with the help of
the twist field $\s _{IJ}$ that corresponds to the group element
$g_{IJ} =1-E_{II}-E_{JJ}+ E_{IJ}+E_{JI}$ transposing the fields $X_I$ and
$X_J$.

The twist fields $\s _g$ have the following OPE
\footnote{Let us stress that there are other primary fields on the r.h.s. of
the OPE, in particular, the field $\s _{g_1g_2}(0)\otimes\bar\s _{g_2g_1}(0)$.
However, these fields will be nonessential in our consideration.}
\be
\s _{g_1}(z,\bar z)\s _{g_2}(0)=\frac{1}{|z|^{2\D _{g_1}+
2\D _{g_2}-2\D _{g_1g_2}}}\left( C_{g_1,g_2}^{g_1g_2}\s _{g_1g_2}(0)
+C_{g_1,g_2}^{g_2g_1}\s _{g_2g_1}(0)\right) +
\cdots .
\label{ope}
\ee
Here the two leading terms appear because there are two different ways to go
around the points $z$ and $0$. It is not difficult to see that $g_1g_2$ and
$g_2g_1$ belong to the same conjugacy class and, hence,
$\D _{g_1g_2}=\D _{g_2g_1}$.

Therefore, the twist field $\s _{IJ}$ acting on the state $\s _g(0)|0\rangle$
creates the states  $\s _{g_{IJ}g}(0)|0\rangle$ and $\s _{gg_{IJ}}(0)|0\rangle$.
An arbitrary element $g$ has a
decomposition $(n_1)(n_2)\cdots (n_k)$ and describes a configuration with $k$
strings. If the indices $I$ and $J$ belong, say, to the cycle $(n_1)$ in the
decomposition then the element $g_{IJ}g$ has the decomposition
$(n_1^{(1)})(n_1^{(2)})(n_2)\cdots (n_k)$ with $n_1^{(1)}+n_1^{(2)}=n_1$ and,
hence, describes a configuration with $k+1$ strings. If the index $I$ belongs
to the cycle $(n_1)$ and the index $J$ belongs to $(n_2)$ then the element
$g_{IJ}g$ has the decomposition $(n_1+n_2)(n_3)\cdots (n_k)$ and
describes a configuration with $k-1$ strings. Thus, the twist field $\s _{IJ}$
generates the elementary joining and splitting of strings.

To write down the DVV interaction vertex it is useful to come back to the
Minkowskian space-time. Then the interaction is described by the
translationally-invariant vertex
\bea
V_{int}=\frac{\lambda N}{2\pi } \sum_{I<J}\int d\tau d\s \s_{IJ}(\s _+,\s _-),
\nonumber
\eea
where $\lambda$ is a coupling constant proportional to the string coupling,
and $\s _\pm$ are light-cone coordinates: $\s _\pm =\tau\pm\s$.

If $D=24$, then the twist field $\s_{IJ}(\s _+,\s _-)$ is a weight
$(\frac {3}{2} ,\frac {3}{2} )$ conformal field and the coupling constant
$\lambda$ has dimension $-1$.

Performing again the Wick rotation and the conformal map onto the sphere,
one gets the following expression for $V_{int}$ (and for $D=24$)
\bea
V_{int}=-\frac{\lambda N}{2\pi } \sum_{I<J}\int d^2z|z| \s_{IJ}(z,\bar z),
\nonumber\eea
where the minus sign appears because $\s_{IJ}$ has conformal dimension
$(\frac {3}{2} ,\frac {3}{2} )$.

Thus, the action of the interacting $S^N\R^{24}$ orbifold sigma model is given
by the sum
\bea
S_{int}=S_0+V_{int}
\nonumber\eea
In the next section we calculate the S-matrix element corresponding to
the scattering of four tachyons and show that the scattering amplitude
coincides with the Virasoro one.

\section{Scattering amplitude}
The S-matrix element at the second order in the coupling constant $\lambda$ is
given by the standard formula of quantum field theory
\be
\langle f|S|i\rangle = -\frac {1}{2}\left(\frac{\lambda N}{2\pi }\right)^2
\langle f|\int d^2z_1d^2z_2|z_1||z_2|T\left(
\cL_{int}(z_1,\bar z_1)
\cL_{int}(z_2,\bar z_2)\right) |i\rangle ,
\label{matel}
\ee
where the symbol $T$ means the time-ordering: $|z_1|>|z_2|$, and
\bea
\cL_{int}(z,\bar z)=\sum_{I<J} \s_{IJ}(z,\bar z).
\nonumber\eea
The initial state $|i\rangle$ describes two tachyons with momenta $\k_1$ and
$\k_2$ and is created by the twist field $\s_{[g_0]}[\k_1,\k_2]$
\bea
|i\rangle =C_0\s_{[g_0]}[\k_1,\k_2](0,0)|0\rangle .
\nonumber\eea
The element $g_0$ is taken in the canonical block-diagonal form
$$g_0=(n_0)(N-n_0),$$
where $n_0<N-n_0$.

The final state $\langle f|$ describes two tachyons with momenta $\k_3$ and
$\k_4$ and is given by the formula (see \cite{BPZ})
\bea
\langle f|=C_\infty \lim_{z_\infty\to\infty}|z_\infty |^{4\D_\infty}
\langle 0|\s_{[g_\infty ]}[\k_3,\k_4](z_\infty ,\bar z_\infty ).
\nonumber\eea
The element $g_\infty$ has the canonical decomposition
$$g_\infty =(n_\infty )(N-n_\infty ),\quad n_\infty <N-n_\infty .$$
The constants $C_0$ and $C_\infty$ are chosen to be equal to
$$
C_0=\sqrt {\frac {N!}{n_0(N-n_0)}},\quad
C_\infty =\sqrt {\frac {N!}{n_\infty (N-n_\infty )}}$$
that guarantees the standard normalization of the initial and final states.

After the conformal transformation $z\to \frac {z}{z_1}$ eq.(\ref{matel})
acquires the form
\bea
\langle f|S|i\rangle &=&-\frac {1}{2}\left(\frac{\lambda N}{2\pi }\right)^2
\int d^2z_1d^2z_2|z_1||z_2||z_1|^{2\D_\infty -2\D_0 -6}\nonumber\\
&\times &\langle f|T\left( \cL_{int}(1,1)
\cL_{int}(\frac {z_2}{z_1},\frac {\bar z_2}{\bar z_1})\right) |i\rangle ,
\label{matel1}
\eea
where $\D_0$ and $\D_\infty$ are conformal dimensions of the twist fields
$\s_{[g_0]}[\k_1,\k_2]$ and $\s_{[g_\infty ]}[\k_3,\k_4]$
\bea
\D_0 &=&N-\frac {1}{n_0}-\frac {1}{N-n_0}+\frac {\k_1^2}{8n_0}+
\frac {\k_2^2}{8(N-n_0)},\nonumber\\
\D_\infty &=&N-\frac {1}{n_\infty}-\frac {1}{N-n_\infty}+\frac {\k_3^2}{8n_\infty}+
\frac {\k_4^2}{8(N-n_\infty )}.
\la{confdim4}
\eea
Let us introduce the light-cone momenta of the tachyons \cite{DVV} taking into
account the mass-shell condition for the tachyonic states
\bea
k_1^+&=&\frac {n_0}{N},\quad k_1^-k_1^+-\k_1^2\equiv -k_1^2 =-8,\nonumber\\
k_2^+&=&\frac {N-n_0}{N},\quad k_2^-k_2^+-\k_2^2\equiv -k_2^2  =-8,\nonumber\\
k_3^+&=&-\frac {n_\infty}{N},\quad k_3^-k_3^+-\k_3^2\equiv -k_3^2  =-8,\nonumber\\
k_4^+&=&-\frac {N-n_\infty}{N},\quad k_4^-k_4^+-\k_4^2\equiv -k_4^2  =-8.
\nonumber\eea
By using the light-cone momenta and the mass-shell condition, one can rewrite
(\ref{confdim4}) in the form
\bea
\D_0 &=&N+\frac {k_1^-+k_2^-}{8N},\nonumber\\
\D_\infty &=&N-\frac {k_3^-+k_4^-}{8N}.
\nonumber\eea
Performing the change of variables $\frac {z_2}{z_1}=u$, one obtains
\bea
\langle f|S|i\rangle &=& -\frac {1}{2}\left(\frac{\lambda N}{2\pi }\right)^2
\int d^2z_1|z_1|^{2\D_\infty -2\D_0 -2}\nonumber\\
&\times& \int d^2u|u|\langle f|T\left( \cL_{int}(1,1)
\cL_{int}(u,\bar u)\right) |i\rangle .
\nonumber\eea
The integral over $z_1$ is obviously divergent. To understand the meaning of
this divergency one should remember that we made the Wick rotation. Coming back
to the $\s ,\tau$-coordinates on the cylinder, we get for the integral over
$z_1$
$$
\int d^2z_1|z_1|^{2\D_\infty -2\D_0 -2}\to
i\int d\tau d\s \e^{2i\tau (\D_\infty -\D_0 )}.
$$
Integration over $\s$ and $\tau$ gives us the conservation law for the
light-cone momenta $k_i^-$
$$
\int d\tau d\s \e^{2i\tau (\D_\infty -\D_0 )}=
4N(2\pi )^2\d (k_1^-+k_2^-+k_3^-+k_4^-).$$
Thus, the S-matrix element is equal to
\bea
\langle f|S|i\rangle =-i2\lambda^2
N^3\d (k_1^-+k_2^-+k_3^-+k_4^-)
\int d^2u|u|\langle f|T\left( \cL_{int}(1,1)
\cL_{int}(u,\bar u)\right) |i\rangle .
\label{matel3}
\eea
So, to find the S-matrix element one has to calculate the correlator
\bea
&&F(u,\bar u)=\langle f|T\left( \cL_{int}(1,1)
\cL_{int}(u,\bar u)\right) |i\rangle \nonumber\\
&&=C_0C_\infty \sum_{I<J;K<L}\langle\s_{[g_\infty ]}[\k_3,\k_4](\infty )T\left(
\s_{IJ}(1,1)\s_{KL}(u,\bar u)\right) \s_{[g_0]}[\k_1,\k_2](0,0)\rangle .
\la{fuu}
\eea
In what follows we assume for definiteness that $n_0<n_\infty$ and $|u|<1$.

By using the definition (\ref{tachop1}) of $\s _{[g]}[\{\k _\a\}]$, and taking
into account that the interaction vertex is $S_N$-invariant, and that any
correlator of twist fields is invariant with respect to the global action of
the symmetric group
\be
\langle \s_{g_1}\s_{g_2}\cdots \s_{g_n}\rangle =
\langle \s_{h^{-1}g_1h}\s_{h^{-1}g_2h}\cdots \s_{h^{-1}g_nh}\rangle ,
\label{invcor}
\ee
we rewrite the correlator in the form
\bea
F(u,\bar u)
=\frac {C_0C_\infty }{N!}\sum_{h_\infty\in S_N}\sum_{I<J;K<L}
\langle\s_{h_\infty ^{-1}g_\infty h_\infty}[\k_3,\k_4](\infty )
\s_{IJ}(1,1)\s_{KL}(u,\bar u)\s_{g_0}[\k_1,\k_2](0,0)\rangle .
\nonumber\eea
Let us note that the correlator
\be
\langle\s_{g_1}(\infty )
\s_{g_2}(1,1)\s_{g_3}(u,\bar u)\s_{g_4}(0,0)\rangle
\la{corr}
\ee
does not vanish only if
\be
g_1g_2g_3g_4=1\quad or\quad g_1g_4g_3g_2=1.
\la{aa}
\ee
It can be seen as follows. Due to the OPE (\ref{ope}) of $\s_g$, in the limit
$u\to 0$ the correlator (\ref{corr}) reduces to the sum of three-point
correlators
$\langle\s_{g_1}\s_{g_2}\s_{g_3g_4}\rangle $ and
$\langle\s_{g_1}\s_{g_2}\s_{g_4g_3}\rangle $. This sum does not vanish if one
of the following equations is fulfilled:
\be
g_1g_2g_3g_4=1,\quad g_1g_3g_4g_2=1,\quad  g_1g_2g_4g_3=1,\quad g_1g_4g_3g_2=1.
\la{aa1}
\ee
From the other side in the limit $u\to 1$ one gets the sum of the correlators
$\langle\s_{g_1}\s_{g_2g_3}\s_{g_4}\rangle $ and
$\langle\s_{g_1}\s_{g_3g_2}\s_{g_4}\rangle $. This sum does not vanish if
\be
g_1g_2g_3g_4=1,\quad g_1g_4g_2g_3=1,\quad  g_1g_3g_2g_4=1,\quad g_1g_4g_3g_2=1.
\la{aa2}
\ee
Comparing eqs.(\ref{aa1}) and (\ref{aa2}), one obtains eq.(\ref{aa}).

However, the contribution of the terms satisfying the equation
$h_\infty ^{-1}g_\infty h_\infty g_0g_{KL}g_{IJ}=1$, coincides with the
contribution of the terms which satisfy
$h_\infty ^{-1}g_\infty h_\infty g_{IJ}g_{KL}g_0=1$. To prove the statement let
us note that the invariance of the action (\ref{act}) with respect to the
world-sheet parity symmetry $z\to \bar z$ (or $\s\to -\s$ in the Minkowskian space-time)
leads to the following equality
\be
\langle \s_{g_1}\s_{g_2}\cdots \s_{g_n}\rangle =
\langle \s_{g_1^{-1}}\s_{g_2^{-1}}\cdots \s_{g_n^{-1}}\rangle ,
\label{invcor1}
\ee
since twist fields $\s_g$ transform into $\s_{g^{-1}}$. Now taking into
account eqs.(\ref{invcor}) and (\ref{invcor1}), and that the elements $g$ and
$g^{-1}$ belong to the same conjugacy class, one obtains the desired equality
\bea
&&\sum_{I<J;K<L}
\langle\s_{g_{IJ}g_{KL}g_0^{-1}}\s_{IJ}\s_{KL}\s_{g_0}\rangle =
\sum_{I<J;K<L}
\langle\s_{g_0g_{KL}g_{IJ}}\s_{IJ}\s_{KL}\s_{g_0^{-1}}\rangle \nonumber\\
&&=\sum_{I<J;K<L}
\langle\s_{g_0^{-1}g_{KL}g_{IJ}}\s_{IJ}\s_{KL}\s_{g_0}\rangle .\nonumber
\eea
Thus, the function $F(u,\bar u)$ is given by a sum of correlators of twist
fields which can be schematically represented as
\bea
{\cal S}=\sum_{h_\infty\in S_N}\sum_{I<J;K<L}
\langle\s_{h_\infty^{-1} g_\infty h_\infty}
\s_{IJ}\s_{KL}\s_{g_0}\rangle ,
\nonumber\eea
where the elements $h_\infty ,g_{IJ},g_{KL}$ solve the equation
$h_\infty ^{-1}g_\infty h_\infty g_{IJ}g_{KL}g_0=1$.

\noindent We can fix the values of the indices $K$ and $L$ by using the action
of the stabilizer of $g_0$ and the invariance (\ref{invcor}) of the correlators
\bea
{\cal S}=\sum_{h_\infty\in S_N}\sum_{I<J}&&\left(
n_0(N-n_0)\langle\s_{h_\infty^{-1} g_\infty h_\infty}
\s_{IJ}\s_{n_0N}\s_{g_0}\rangle \right.\nonumber\\
&&+\left. (N-n_0)\langle\s_{h_\infty^{-1} g_\infty h_\infty}
\s_{IJ}\s_{n_\infty N}\s_{g_0}\rangle \right.\nonumber\\
&&+\left. (N-n_0)\langle\s_{h_\infty^{-1} g_\infty h_\infty}
\s_{IJ}\s_{n_0+n_\infty ,N}\s_{g_0}\rangle \right) .
\la{s1}
\eea
The first term in (\ref{s1}) corresponds to the joining of two incoming strings
and the factor $n_0(N-n_0)$ appears since in this case the index $K$ can take
$n_0$ values, $K=1,...,n_0$, and the index $L$ takes $N-n_0$ values,
$L=n_0+1,...,N$. To fix $K=n_0$ and $L=N$ we have to use all elements of
$C_{g_0}$. The second and the third terms correspond to the splitting of the
string of length $N-n_0$ into two strings of lengths $n_\infty -n_0$ and
$N-n_\infty$, and $N-n_0-n_\infty$ and $n_\infty$ respectively. In these cases
to fix the values of $K$ and $L$ one should use $N-n_0$ elements of the
subgroup $\Z_{N-n_0}$ of $C_{g_0}$ that does not act on the cycle $(n_0)$.

Eq.(\ref{s1}) can be further rewritten in the form
\bea
{\cal S}&=&n_0(N-n_0)n_\infty (N-n_\infty )
\left(\sum_{I=1}^{n_\infty}
\langle\s_{g_\infty (I)}
\s_{I,I+N-n_\infty}\s_{n_0N}\s_{g_0}\rangle \right.\nonumber\\
&+&\sum_{I=1}^{N-n_\infty}\langle\s_{g_\infty (I)}
\s_{I,I+n_\infty}\s_{n_0 N}\s_{g_0}\rangle +
\sum_{J=n_0+1}^{n_\infty}
\langle\s_{g_\infty (J)}
\s_{n_0J}\s_{n_\infty N}\s_{g_0}\rangle
\nonumber\\
&+&\left.\sum_{J=n_0+n_\infty +1}^{N}
\langle\s_{g_\infty (J)}
\s_{n_0J}\s_{n_0+n_\infty ,N}\s_{g_0}
\rangle \right),
\la{s2}
\eea
where the elements $g_\infty $ have to be found from the equation $g_\infty
g_{IJ}g_{KL}g_0=1$.

Some comments are in order. The factor $n_\infty (N-n_\infty )$ is the volume
of the stabilizer $\Z_{n_\infty}\times\Z_{N-n_\infty}$ of $g_\infty$. The first
two terms correspond to the splitting of the long string of length $N$ into
strings of lengths $n_\infty$ and $N-n_\infty$. It can be achieved only if
$J-I=N-n_\infty$ or $J-I=n_\infty$. In the third and fourth terms we fixed the
value of $I$ equal to $n_0$ by using the action of the subgroup $\Z_{n_0}$ of
$C_{g_0}$. It gave the additional factor $n_0$. The third (fourth) term
describes the joining of the strings of lengths $n_0$ and $n_\infty -n_0$
($N-n_0-n_\infty $) into one string of length $n_\infty$ ($N-n_\infty $). The
total number of different correlators is, therefore, equal to $2(N-n_0)$.
The  diagramms corresponding to these four terms are depicted in Fig.1.
\begin{figure}[t]
\special{em:linewidth 0.4pt}
\unitlength 1.00mm
\linethickness{0.4pt}
\begin{picture}(130.33,145.62)(-10,0)
\put(13.67,145.00){\circle{13.23}}
\put(13.67,132.00){\circle{4.06}}
\put(140.67,132.00){\circle{6.32}}
\put(140.67,145.00){\circle{8.67}}
\put(13.67,105.00){\circle{12.81}}
\put(13.67,90.00){\circle{4.81}}
\put(13.33,65.00){\circle{11.66}}
\put(13.67,50.00){\circle{4.81}}
\put(13.67,25.00){\circle{11.33}}
\put(13.67,10.00){\circle{4.85}}
\put(141.00,90.00){\circle{8.03}}
\put(141.33,105.00){\circle{4.71}}
\put(141.33,65.00){\circle{8.67}}
\put(141.33,50.00){\circle{6.04}}
\put(142.00,10.00){\circle{8.00}}
\put(142.00,25.00){\circle{6.00}}
\emline{15.00}{151.33}{1}{38.92}{149.91}{2}
\emline{38.92}{149.91}{3}{62.76}{148.95}{4}
\emline{62.76}{148.95}{5}{86.54}{148.45}{6}
\emline{86.54}{148.45}{7}{114.18}{148.44}{8}
\emline{114.18}{148.44}{9}{140.00}{149.00}{10}
\emline{14.67}{138.33}{11}{30.66}{138.19}{12}
\emline{30.66}{138.19}{13}{43.17}{137.85}{14}
\emline{43.17}{137.85}{15}{54.94}{137.32}{16}
\emline{54.94}{137.32}{17}{60.33}{137.00}{18}
\emline{14.33}{133.67}{19}{59.67}{137.00}{20}
\emline{14.33}{130.33}{21}{40.03}{131.10}{22}
\emline{40.03}{131.10}{23}{64.87}{131.40}{24}
\emline{64.87}{131.40}{25}{84.91}{131.28}{26}
\emline{84.91}{131.28}{27}{104.36}{130.82}{28}
\emline{104.36}{130.82}{29}{123.21}{130.04}{30}
\emline{123.21}{130.04}{31}{140.33}{129.00}{32}
\emline{140.33}{140.67}{33}{90.67}{139.33}{34}
\emline{140.33}{134.67}{35}{131.48}{135.18}{36}
\emline{131.48}{135.18}{37}{119.64}{136.08}{38}
\emline{119.64}{136.08}{39}{106.93}{137.29}{40}
\emline{106.93}{137.29}{41}{91.67}{139.00}{42}
\emline{15.67}{111.00}{43}{42.23}{109.33}{44}
\emline{42.23}{109.33}{45}{69.46}{108.15}{46}
\emline{69.46}{108.15}{47}{97.39}{107.45}{48}
\emline{97.39}{107.45}{49}{130.13}{107.25}{50}
\emline{130.13}{107.25}{51}{141.33}{107.33}{52}
\emline{14.00}{87.67}{53}{49.05}{88.08}{54}
\emline{49.05}{88.08}{55}{84.79}{87.79}{56}
\emline{84.79}{87.79}{57}{121.23}{86.81}{58}
\emline{121.23}{86.81}{59}{141.00}{86.00}{60}
\emline{14.00}{98.67}{61}{60.00}{96.00}{62}
\emline{14.00}{92.33}{63}{19.96}{92.62}{64}
\emline{19.96}{92.62}{65}{28.84}{93.20}{66}
\emline{28.84}{93.20}{67}{44.12}{94.51}{68}
\emline{44.12}{94.51}{69}{58.67}{96.00}{70}
\emline{141.67}{102.67}{71}{90.00}{99.67}{72}
\emline{91.67}{99.67}{73}{93.51}{99.17}{74}
\emline{93.51}{99.17}{75}{95.41}{98.69}{76}
\emline{95.41}{98.69}{77}{97.39}{98.23}{78}
\emline{97.39}{98.23}{79}{99.43}{97.80}{80}
\emline{99.43}{97.80}{81}{101.55}{97.38}{82}
\emline{101.55}{97.38}{83}{103.73}{96.98}{84}
\emline{103.73}{96.98}{85}{105.99}{96.61}{86}
\emline{105.99}{96.61}{87}{108.31}{96.25}{88}
\emline{108.31}{96.25}{89}{110.70}{95.92}{90}
\emline{110.70}{95.92}{91}{113.16}{95.61}{92}
\emline{113.16}{95.61}{93}{115.69}{95.32}{94}
\emline{115.69}{95.32}{95}{118.29}{95.05}{96}
\emline{118.29}{95.05}{97}{123.70}{94.57}{98}
\emline{123.70}{94.57}{99}{129.39}{94.17}{100}
\emline{129.39}{94.17}{101}{135.36}{93.86}{102}
\emline{135.36}{93.86}{103}{140.33}{93.67}{104}
\emline{14.00}{70.33}{105}{24.92}{69.41}{106}
\emline{24.92}{69.41}{107}{36.06}{68.65}{108}
\emline{36.06}{68.65}{109}{47.43}{68.06}{110}
\emline{47.43}{68.06}{111}{62.92}{67.53}{112}
\emline{62.92}{67.53}{113}{78.80}{67.30}{114}
\emline{78.80}{67.30}{115}{95.07}{67.36}{116}
\emline{95.07}{67.36}{117}{111.73}{67.71}{118}
\emline{111.73}{67.71}{119}{128.78}{68.37}{120}
\emline{128.78}{68.37}{121}{141.00}{69.00}{122}
\emline{13.67}{47.67}{123}{34.06}{48.62}{124}
\emline{34.06}{48.62}{125}{54.32}{49.18}{126}
\emline{54.32}{49.18}{127}{74.45}{49.35}{128}
\emline{74.45}{49.35}{129}{94.45}{49.12}{130}
\emline{94.45}{49.12}{131}{114.32}{48.50}{132}
\emline{114.32}{48.50}{133}{130.12}{47.73}{134}
\emline{130.12}{47.73}{135}{141.67}{47.00}{136}
\emline{13.67}{59.33}{137}{69.51}{56.12}{138}
\emline{69.51}{56.12}{139}{89.67}{55.33}{140}
\emline{13.67}{52.33}{141}{88.67}{55.33}{142}
\emline{141.00}{52.67}{143}{107.00}{56.24}{144}
\emline{107.00}{56.24}{145}{77.03}{59.96}{146}
\emline{77.03}{59.96}{147}{67.00}{61.33}{148}
\emline{68.00}{61.00}{149}{118.93}{60.56}{150}
\emline{118.93}{60.56}{151}{141.33}{60.67}{152}
\emline{13.67}{30.67}{153}{32.06}{29.38}{154}
\emline{32.06}{29.38}{155}{51.05}{28.41}{156}
\emline{51.05}{28.41}{157}{70.64}{27.76}{158}
\emline{70.64}{27.76}{159}{94.94}{27.41}{160}
\emline{94.94}{27.41}{161}{120.10}{27.53}{162}
\emline{120.10}{27.53}{163}{142.00}{28.00}{164}
\emline{13.67}{7.67}{165}{23.79}{8.22}{166}
\emline{23.79}{8.22}{167}{37.89}{8.73}{168}
\emline{37.89}{8.73}{169}{52.70}{8.97}{170}
\emline{52.70}{8.97}{171}{68.22}{8.96}{172}
\emline{68.22}{8.96}{173}{88.61}{8.58}{174}
\emline{88.61}{8.58}{175}{114.53}{7.59}{176}
\emline{114.53}{7.59}{177}{142.33}{6.00}{178}
\emline{14.00}{19.33}{179}{85.00}{15.33}{180}
\emline{14.00}{12.33}{181}{41.68}{13.81}{182}
\emline{41.68}{13.81}{183}{84.00}{15.33}{184}
\emline{67.00}{21.67}{185}{97.00}{18.07}{186}
\emline{97.00}{18.07}{187}{131.15}{14.63}{188}
\emline{131.15}{14.63}{189}{142.00}{13.67}{190}
\emline{67.33}{21.67}{191}{119.89}{21.69}{192}
\emline{119.89}{21.69}{193}{142.00}{22.00}{194}
\put(146.00,132.00){\makebox(0,0)[lc]{$n_{\infty},{\bf k}_3$}}
\put(146.33,145.00){\makebox(0,0)[lc]{$N-n_{\infty},{\bf k}_4$}}
\put(5.33,145.00){\makebox(0,0)[rc]{$N-n_0,{\bf k}_2$}}
\put(9.67,132.00){\makebox(0,0)[rc]{$n_0,{\bf k}_1$}}
\put(5.33,145.00){\makebox(0,0)[rc]{$N-n_0,{\bf k}_2$}}
\put(6.00,105.00){\makebox(0,0)[rc]{$N-n_0,{\bf k}_2$}}
\put(10.33,90.00){\makebox(0,0)[rc]{$n_0,{\bf k}_1$}}
\put(6.33,64.67){\makebox(0,0)[rc]{$N-n_0,{\bf k}_2$}}
\put(10.00,49.67){\makebox(0,0)[rb]{$n_0,{\bf k}_1$}}
\put(7.00,24.67){\makebox(0,0)[rc]{$N-n_0,{\bf k}_2$}}
\put(10.33,9.67){\makebox(0,0)[rc]{$n_0,{\bf k}_1$}}
\put(145.00,104.67){\makebox(0,0)[lc]{$n_{\infty},{\bf k}_3$}}
\put(146.67,90.00){\makebox(0,0)[lc]{$N-n_{\infty},{\bf k}_4$}}
\put(147.00,65.00){\makebox(0,0)[lc]{$N-n_{\infty},{\bf k}_4$}}
\put(145.67,50.00){\makebox(0,0)[lc]{$n_{\infty},{\bf k}_3$}}
\put(146.67,25.00){\makebox(0,0)[lc]{$n_{\infty},{\bf k}_3$}}
\put(147.33,10.00){\makebox(0,0)[lc]{$N-n{\infty},{\bf k}_4$}}
\end{picture}
\caption{The diagramm representation of different correlators in
eq.(4.32)}
\label{fig1}
\end{figure}
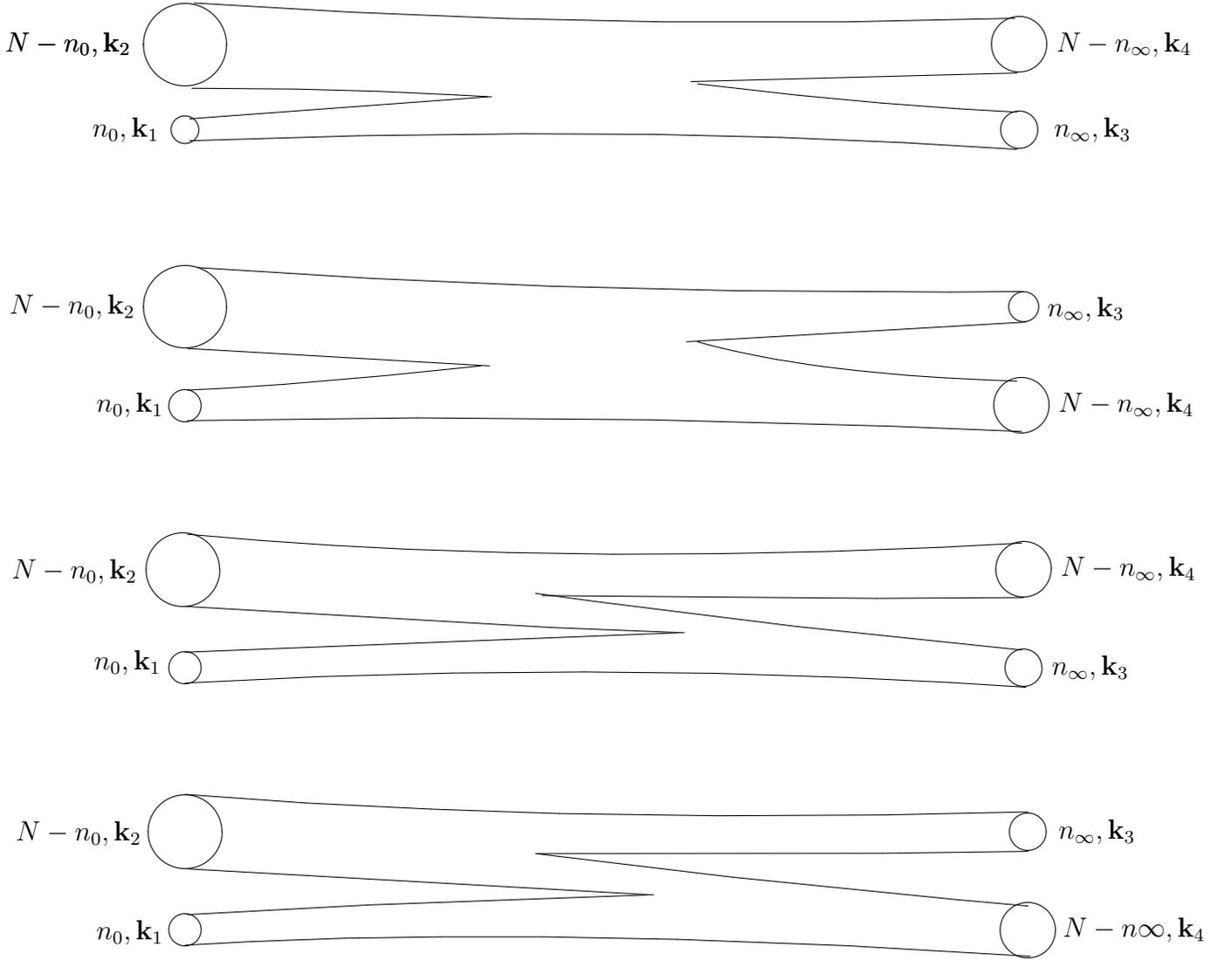

So, we need to compute the correlators (and the same correlators with the
interchange $u\leftrightarrow 1$)
\be
G(u,\bar u)=
\langle\s_{ g_\infty}[\k_3,\k_4](\infty )
\s_{IJ}(1,1)\s_{KL}(u,\bar u) \s_{g_0}[\k_1,\k_2](0,0)\rangle ,
\la{guu}
\ee
where all possible elements $g_\infty ,g_{IJ},g_{KL},g_0$ are listed in
eq.(\ref{s2}).

To calculate the correlator (\ref{guu}) we employ the stress-energy tensor
method \cite{DFMS}. The idea of the method is as follows. Let us suppose that we
know the following ratio
\be
f(z,u)=
\frac {\langle T(z)\phi_\infty (\infty )\phi_1(1)\phi_2(u)\phi_0(0)\rangle }
{\langle \phi_\infty (\infty )\phi_1(1)\phi_2(u)\phi_0(0)\rangle },
\la{ratio}
\ee
where $T(z)$ is the stress-energy tensor and $\phi$ are primary fields. Taking
into account that the OPE of $T(z)$ with any primary field has the form
\bea
T(z)\phi (0)=\frac {\D}{z^2}\phi (0) +\frac {1}{z}\partial\phi (0)+\cdots ,
\nonumber
\eea
one gets a differential equation on the correlator
$G(u,\bar u)=\langle \phi_\infty (\infty )\phi_1(1)\phi_2(u)\phi_0(0)\rangle$
\bea
\partial _u\log G(u,\bar u)=H(u,\bar u),
\nonumber\eea
where $H(u,\bar u)$ is the second term in the decomposition of the function
$f(z,u)$ in the vicinity of $u$
\bea
f(z,u)=\frac {\D_2}{(z-u)^2} +\frac {1}{z-u}H(u,\bar u)+\cdots .
\nonumber
\eea
In the same way one gets the second equation on $G(u,\bar u)$ by using
the stress-energy tensor $\bar T(\bar z)$
\bea
\partial _{\bar u}\log G(u,\bar u)=\bar H(u,\bar u).
\nonumber\eea
A solution of these two equations determines the correlator
$G(u,\bar u)$ up to a constant.

To calculate the ratio (\ref{ratio}) we firstly find the following Green
functions
\footnote{We consider the correlators for general values of $D$ keeping in mind
the application to the superstring case.}
\bea
G^{ij}_{MS}(z,w)
&=&
\frac {\langle \partial X^i_M(z)\partial X^j_S(w)\s_{ g_\infty}[\k_3,\k_4](\infty )
\s_{IJ}(1,1)\s_{KL}(u,\bar u) \s_{g_0}[\k_1,\k_2](0,0)\rangle }
{\langle \s_{ g_\infty}[\k_3,\k_4](\infty )
\s_{IJ}(1,1)\s_{KL}(u,\bar u) \s_{g_0}[\k_1,\k_2](0,0)\rangle }\nonumber\\
&\equiv&\langle\langle \partial X^i_M(z)\partial X^j_S(w)\rangle\rangle  .
\nonumber\eea
These Green functions have non-trivial monodromies around points
$\infty ,1,u$ and $0$, and, in fact, are different branches of
one multi-valued function. However, this function is single-valued on the
sphere that is obtained by gluing the fields $X^i_I$ at $z=0$ and $z=\infty$.
Thus to construct $G^{ij}_{MS}(z,w)$ we introduce the following
map from this sphere onto the original one:
\be
z=\frac {t^{n_0}(t-t_0)^{N-n_0}}{(t-t_\infty )^{N-n_\infty}}
\frac {(t_1-t_\infty )^{N-n_\infty}}{t_1^{n_0}(t_1-t_0)^{N-n_0}}
\equiv u(t).
\la{map}
\ee
Here the points $t=0$ and $t=t_0$ are mapped  to the point $z=0$;
$t=\infty $, $t=t_\infty\to z=\infty$,
$t=t_1\to z=1$ and $t=x\to z=u$ (see Fig.2).
\begin{figure}
\special{em:linewidth 0.4pt}
\unitlength 1.00mm
\linethickness{0.4pt}
\begin{picture}(128.67,67.67)(0,30)
\emline{27.00}{34.67}{1}{52.67}{60.33}{2}
\emline{52.67}{60.33}{3}{128.67}{60.33}{4}
\emline{128.67}{60.33}{5}{106.67}{35.00}{6}
\emline{106.67}{35.00}{7}{27.33}{34.67}{8}
\emline{27.00}{74.00}{9}{52.67}{97.67}{10}
\emline{52.67}{97.67}{11}{128.00}{97.67}{12}
\emline{128.00}{97.67}{13}{107.00}{73.67}{14}
\emline{107.00}{73.67}{15}{27.67}{74.33}{16}
\put(57.67,46.00){\vector(1,-4){0.2}}
\emline{49.33}{82.00}{17}{57.67}{46.00}{18}
\put(103.33,46.00){\vector(1,-4){0.2}}
\emline{97.67}{81.33}{19}{103.33}{46.00}{20}
\put(103.33,46.00){\vector(-1,-4){0.2}}
\emline{111.67}{90.33}{21}{103.33}{46.00}{22}
\put(76.00,41.67){\vector(0,-1){0.2}}
\emline{75.00}{83.00}{23}{76.00}{41.67}{24}
\put(89.67,53.00){\vector(0,-1){0.2}}
\emline{89.00}{87.33}{25}{89.67}{53.00}{26}
\put(75.00,86.33){\makebox(0,0)[cc]{$x$}}
\put(89.00,90.33){\makebox(0,0)[cc]{$t_1$}}
\put(97.67,84.00){\makebox(0,0)[cc]{$t_{\infty}$}}
\put(112.00,92.67){\makebox(0,0)[cc]{$\infty$}}
\put(37.00,77.67){\makebox(0,0)[cc]{$t$}}
\put(37.00,38.67){\makebox(0,0)[cc]{z}}
\put(49.33,84.33){\makebox(0,0)[cc]{$0$}}
\put(57.67,41.00){\makebox(0,0)[cc]{$0$}}
\put(75.67,37.67){\makebox(0,0)[cc]{$u$}}
\put(103.00,41.67){\makebox(0,0)[cc]{$\infty$}}
\put(49.67,81.67){\rule{0.00\unitlength}{0.00\unitlength}}
\put(63.33,90.33){\rule{0.00\unitlength}{0.00\unitlength}}
\put(75.33,82.67){\rule{0.00\unitlength}{0.00\unitlength}}
\put(88.67,87.33){\rule{0.00\unitlength}{0.00\unitlength}}
\put(97.67,81.00){\rule{0.00\unitlength}{0.00\unitlength}}
\put(111.67,90.00){\rule{0.00\unitlength}{0.00\unitlength}}
\put(63.67,93.67){\makebox(0,0)[cc]{$t_0$}}
\put(89.67,49.00){\makebox(0,0)[cc]{$1$}}
\put(49.33,81.67){\circle*{0.67}}
\put(63.67,90.33){\circle*{1.49}}
\put(49.33,81.67){\circle*{1.49}}
\put(75.00,82.67){\circle*{1.33}}
\put(89.00,87.00){\circle*{0.00}}
\put(89.00,87.00){\circle*{1.33}}
\put(98.00,80.67){\circle*{1.33}}
\put(111.67,90.00){\circle*{1.49}}
\put(57.67,46.00){\vector(-1,-4){0.2}}
\emline{63.67}{90.00}{27}{57.67}{46.00}{28}
\put(57.67,44.00){\circle*{1.33}}
\put(76.00,41.00){\circle*{1.49}}
\put(89.67,51.67){\circle*{1.33}}
\put(103.00,44.33){\circle*{1.49}}
\end{picture}
\label{fig2}
\caption{The $N$-fold covering of the $z$-sphere by the $t$-sphere.}
\end{figure}
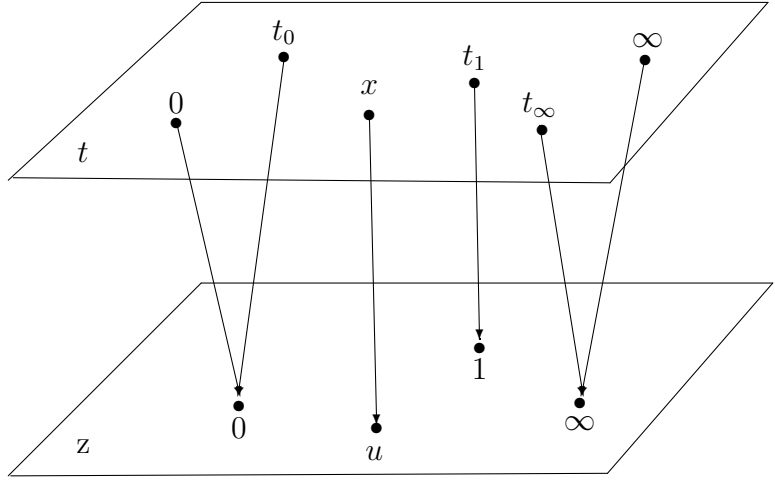
The map (\ref{map}) may be viewed as the $N$-fold covering of the
$z$-plain by the $t$-sphere on which the Green function  is
single-valued. The more detailed discussion of eq.(\ref{map}) is
presented in the Appendix.

Due to the projective transformations, the positions of
the points $t_0,t_\infty ,t_1$ depend on $x$ and it is convinient to choose
this dependence as follows
\bea
t_0&=&x-1,\nonumber\\
t_\infty &=&x-\frac {(N-n_\infty )x}{(N-n_0)x+n_0},\nonumber\\
t_1&=&\frac {N-n_0-n_\infty }{n_\infty}+\frac {n_0x}{n_\infty}-
\frac {N(N-n_\infty )x}{n_\infty ((N-n_0)x+n_0)}.
\nonumber\eea
This choice leads to the following expression for
the rational function $u(x)$
\bea
u=u(x)&=&(n_0-n_{\infty})^{n_0-n_{\infty}}\frac{n_{\infty}^{n_{\infty}}}
{n_0^{n_0}}\left(\frac{N-n_0}{N-n_{\infty}}\right)^{N-n_{\infty}}
\left(\frac{x+\frac{n_0}{N-n_0}}{x-1}\right)^N\nonumber\\
&\times&\left(\frac{x-\frac{N-n_0-n_{\infty}}{N-n_0}}{x}\right)^{N-n_0-n_{\infty}}
\left(x-\frac{n_0}{n_0-n_{\infty}}\right)^{n_0-n_{\infty}}.
\la{ux}
\eea
Since $n_0<n_{\infty}$, the map $u(x)$ can be treated as the
$2(N-n_0)$-fold covering of the $u$-sphere by the $x$-sphere, that
means that an equation $u(x)=u$ has $2(N-n_0)$ different
solutions. It is worthwhile to note that
this number coincides with the number of nontrivial correlators in eq.(\ref{s2})
and, therefore different roots of eq.(\ref{ux}) correspond to different
correlators (\ref{s2}). We see that the $t$-sphere can be represented as the
union of $2(N-n_0)$ domains, and each domain $V_{IJKL}$ contains the points $x$
corresponding to the correlator (\ref{guu}). If we take on the
$u$-plain the appropriate system of cuts, then every root of eq.(\ref{ux})
realizes a one-to-one conformal mapping of the cut $u$-plain onto
the corresponding domain $V_{IJKL}$.

Let us now choose some root of eq.(\ref{ux}).
One can always cut the $z$-sphere and numerate
the roots $t_R(z)$ of eq.(\ref{map}) in such a way that they have the
same monodromies as the fields $X$ do.
Then the Green functions are
obviously not equal to zero only if $\k_1+\k_2+\k_3+\k_4=0$ and are
given by
\bea G_{MS}^{ij}(z,w)&=& -\d^{ij}\frac
{t_M'(z)t_S'(w)}{(t_M(z)-t_S(w))^{2}} -\frac
{k_1^ik_1^jt_M'(z)t_S'(w)}{4t_M(z)t_S(w)}\nonumber\\ &-&\frac
{k_1^ik_2^jt_M'(z)t_S'(w)}{4t_M(z)(t_S(w)-t_0)} -\frac
{k_2^ik_1^jt_M'(z)t_S'(w)}{4(t_M(z)-t_0)t_S(w)}\nonumber\\ &-&\frac
{k_2^ik_2^jt_M'(z)t_S'(w)}{4(t_M(z)-t_0)(t_S(w)-t_0)} -\frac
{k_1^ik_4^jt_M'(z)t_S'(w)}{4t_M(z)(t_S(w)-t_\infty )}\nonumber\\
&-&\frac {k_2^ik_4^jt_M'(z)t_S'(w)}{4(t_M(z)-t_0)(t_S(w)-t_\infty )}
-\frac {k_4^ik_1^jt_M'(z)t_S'(w)}{4(t_M(z)-t_\infty )t_S(w)}\nonumber\\
&-&\frac {k_4^ik_2^jt_M'(z)t_S'(w)}{4(t_M(z)-t_\infty )(t_S(w)-t_0)}
-\frac {k_4^ik_4^jt_M'(z)t_S'(w)}{4(t_M(z)-t_\infty )(t_S(w)-t_\infty )}.
\la{greenf1}
\eea
One can easily check that these functions have the singularity
$-\frac {\d^{ij}\d_{MS}}{(z-w)^{2}}$ in the vicinity $z-w=0$ and proper
monodromies at points $z=\infty ,1,u,0$.

Recall that the stress-energy tensor is defined as
\bea
T(z)= -\frac 12 \lim _{w\to z}\sum_{i=1}^{D}\sum_{I=1}^N\left(
\partial X_I^i(z)\partial X_I^i(w) +\frac {1}{(z-w)^2}\right).
\nonumber
\eea
By using this definition and eq.(\ref{greenf1}), one gets
\footnote{If all $\k_i=0$, the expectation value of
$T(z)$ in the presence of twist fields can be equivalently found by
mapping with $t_M(z)$ the stress-energy tensor on the $t$-sphere onto
the $z$-sphere with the subsequent summation over $M$ (see
e.g.\cite{DFMS})}
\bea
\langle\langle T(z)\rangle\rangle &=&
\sum_M\left(\frac {D}{12}\left(\left(\frac {t_M''(z)}{t_M'(z)}\right)'-
\frac {1}{2}\left(\frac {t_M''(z)}{t_M'(z)}\right)^2\right)
+\frac {\k_1^2(t_M'(z))^2}{8(t_M(z))^2}\right.\nonumber\\
&+&\frac {\k_1\k_2(t_M'(z))^2}{4t_M(z)(t_M(z)-t_0)}
+\frac {\k_2^2(t_M'(z))^2}{8(t_M(z)-t_0)^2}\nonumber\\
&+&\frac {\k_1\k_4(t_M'(z))^2}{4t_M(z)(t_M(z)-t_\infty )}
+\frac {\k_2\k_4(t_M'(z))^2}{4(t_M(z)-t_0)(t_M(z)-t_\infty )}\nonumber\\
&+&\left.\frac {\k_4^2(t_M'(z))^2}{8(t_M(z)-t_\infty )^2}\right) .
\nonumber\eea
The term
$$\left(\frac {t''}{t'}\right) '-
\frac {1}{2}\left(\frac {t''}{t'}\right)^2=
\frac {t'''}{t'}-\frac {3}{2}\left(\frac {t''}{t'}\right)^2$$
is the Schwartz derivative as one could expect from the very beginning. To get
the differential equation on the correlator (\ref{guu}) one should expand
$\langle\langle T(z)\rangle\rangle $ in the vicinity of $z=u$. This expansion
is given by
\bea
\langle\langle T(z)\rangle\rangle &=&\frac {D}{16(z-u)^2}-\frac {D}{16(z-u)u}\left( 1+
\frac {2a_2}{a_0^{2}}-\frac {3a_1^2}{2a_0^3}\right) \nonumber\\
&+&
\frac {1}{4a_0(z-u)u}\left( \frac {\k_1^2}{x^2}
+ \frac {\k_2^2}{(x-t_0)^2}+
\frac {2\k_1\k_2}{x(x-t_0)}+
\frac {\k_4^2}{(x-t_\infty )^2}\right.\nonumber\\
&+& \left.\frac {2\k_1\k_4}{x(x-t_\infty )}+
\frac {2\k_2\k_4}{(x-t_0)(x-t_\infty )}\right) +\cdots .
\la{dect}
\eea
Here the coefficients $a_k$ are defined as follows
\bea
a_k=\frac {(-1)^{k-1}}{k+2}\left( \frac {n_0}{x^{k+2}}+
\frac {N-n_0}{(x-t_0)^{k+2}}-\frac {N-n_\infty}{(x-t_\infty)^{k+2}}\right) .
\nonumber
\eea
The first term shows that the conformal dimension of the twist field $\s_{KL}$
is equal to
$\frac {D}{16}$, as it should be, and the other terms lead to the following
differential
equation on $G(u,\bar u)$
\bea
u\partial_u \log G(u,\bar u)&=&-\frac {D}{16}\left( 1+
\frac {2a_2}{a_0^{2}}-\frac {3a_1^2}{2a_0^3}\right) \nonumber\\
&+&
\frac {1}{4a_0}\left( \frac {\k_1^2}{x^2}
+ \frac {\k_2^2}{(x-t_0)^2}+
\frac {2\k_1\k_2}{x(x-t_0)}+
\frac {\k_4^2}{(x-t_\infty )^2}\right.\nonumber\\
&+& \left.\frac {2\k_1\k_4}{x(x-t_\infty )}+
\frac {2\k_2\k_4}{(x-t_0)(x-t_\infty )}\right) .
\la{difur3}
\eea
It is useful to make the change of variables $u\to u(x)$. Then, performing
simple but tedious calculations which are outlined in Appendix, one obtains the
following differential equation on $G(u,\bar u)$
\bea
\partial_x \log G(u(x),\bar u(\bar x)) &=&
-\frac {D}{16}\frac {d}{dx}\log u
+\frac {d_0}{x}
+\frac {d_1}{x-1}
+\frac {d_2}{x+\frac {n_0}{N-n_0}}\nonumber\\
&+&\frac {d_3}{x-\frac {N-n_0-n_\infty }{N-n_0}}
+\frac {d_4}{x-\frac {n_0}{n_0-n_\infty }}
-\frac {D}{24}(\frac {1}{x-\alpha_1}+\frac {1}{x-\alpha_2}) .
\la{difur4}
\eea
Here
$$
\alpha_i =\frac {n_0}{n_0-n_\infty }
+(-1)^i \sqrt {\frac {n_0n_\infty (N-n_\infty)}{(n_0-n_\infty )^2(N-n_0)}}
$$
are roots of the equation $x^2a_0=0$ and the coefficients $d_i$ are given by
the following formulas
\bea
d_0&=& \frac {D}{24}+\frac {n_0}{8(N-n_\infty )}(\k_4^2-\frac {D}{3})+
\frac {N-n_\infty }{8n_0}(\k_1^2-\frac {D}{3})+\frac {1}{4}\k_1\k_4 ,
\nonumber\\
d_1&=& \frac {D}{24}-\frac {n_\infty }{8(N-n_\infty )}(\k_4^2-\frac {D}{3})-
\frac {N-n_\infty }{8n_\infty }(\k_3^2-\frac {D}{3})+\frac {1}{4}\k_3\k_4 ,
\nonumber\\
d_2&=& \frac {D}{24}-\frac {n_0}{8(N-n_0 )}(\k_2^2-\frac {D}{3})-
\frac {N-n_0}{8n_0}(\k_1^2-\frac {D}{3})+\frac {1}{4}\k_1\k_2 ,
\nonumber\\
d_3&=& \frac {D}{24}+\frac {n_\infty }{8(N-n_0)}(\k_2^2-\frac {D}{3})+
\frac {N-n_0 }{8n_\infty}(\k_3^2-\frac {D}{3})+\frac {1}{4}\k_2\k_3 ,\nonumber\\
d_4&=& \frac {D}{24}+\frac {n_0}{8n_\infty }(\k_3^2-\frac {D}{3})+
\frac {n_\infty }{8n_0}(\k_1^2-\frac {D}{3})+\frac {1}{4}\k_1\k_3 .
\la{defd}
\eea
Taking into account that the second equation on $G(u,\bar u)$ has the same form
with the obvious
substitution $u\to\bar u,x\to\bar x $, one gets the
solution of eq.(\ref{difur4})
\bea
G(u,\bar u)&=&C(g_0,g_\infty )\d_R^D(\k_1+\k_2+\k_3+\k_4)|u|^{-\frac {D}{8}}
|x-\alpha_1|^{-\frac {D}{12}}|x-\alpha_2|^{-\frac {D}{12}}
\nonumber\\
&\times& |x|^{2d_0}|x-1|^{2d_1}|x+\frac {n_0}{N-n_0}|^{2d_2}
|x-\frac {N-n_0-n_\infty }{N-n_0}|^{2d_3}
|x-\frac {n_0}{n_0-n_\infty }|^{2d_4}.
\la{soldif}
\eea
Here $x=x(u)$ is the root of equation $u=u(x)$ that corresponds to given values
of the indices $I,J,K,L$, and $C(g_0,g_\infty )$ is a normalization constant
which does not depend on $u,\bar u$.

To determine this constant let us consider an auxiliary correlator
\be
G_0(u,\bar u)=
\langle\s_{ g_0^{-1}}[-\k_1,-\k_2](\infty )
\s_{IJ}(1,1)\s_{IJ}(u,\bar u) \s_{g_0}[\k_1,\k_2](0,0)\rangle ,
\la{gouu}
\ee
where $I=1,\ldots ,n_0$, $J=n_0+1,\ldots ,N$.

\noindent Let us note that by using the action of $C_{g_0}$ one can fix
$I=n_0$, $J=N$. This correlator corresponds to the case $n_\infty =n_0$ and the
rational function $u(x)$ is equal to
\be
u(x)=\left( 1+\frac {2n_0-N}{N-n_0}\frac {1}{x}\right)^{N-2n_0}
\left(\frac { 1+\frac {n_0}{N-n_0}\frac {1}{x}}{1-\frac {1}{x}}
\right)^{N}.
\la{uxo}
\ee
The root of eq.(\ref{uxo}) that corresponds to the correlator (\ref{gouu})
behaves as
\be
\frac {1}{x}=\frac {1}{4n_0}(u-1) +o(u-1),\qquad \mbox{ when} \quad u\to 1.
\la{asym}
\ee
The following expression for the correlator $G_0(u,\bar u)$ can be derived from
eq.(\ref{soldif}) in the limit $n_\infty\to n_0$
\bea
G_0(u,\bar u)&=&C(g_0)R^D|u|^{-\frac {D}{8}}
|x-\frac {N-2n_0}{2(N-n_0)}|^{-\frac {D}{12}}\nonumber\\
&\times& |x|^{2d_0}|x-1|^{2d_1}|x+\frac {n_0}{N-n_0}|^{2d_2}
|x-\frac {N-2n_0}{N-n_0}|^{2d_3},
\la{soldif1}
\eea
where the coefficients $d_i$ are given by eq.(\ref{defd}) with the obvious
substitution $n_\infty\to n_0$, $\k_3=-\k_1$ and $\k_4=-\k_2$.

Taking into account the OPE
\bea
\s_{IJ}(1,1)\s_{IJ}(u,\bar u)=\frac {R^{-D}}{|u-1|^{\frac
{D}{4}}}+\cdots ,
\nonumber
\eea
and the normalization (\ref{norm}) of two-point correlation functions, one gets
\be
G_0(u,\bar u)\rightarrow \frac {R^{D}}{|u-1|^{\frac {D}{4}}}.
\la{asgo1}
\ee
From the other side by using eqs.(\ref{asym}) and (\ref{soldif1}), one derives
 in the limit $u\to 1$
\be
G_0(u,\bar u)\rightarrow C(g_0)R^D
\left( \frac {1}{4n_0}|u-1|\right)^{-2(d_0+d_1+d_2+d_3-\frac
{D}{24})}
=\frac {R^{D}}{|u-1|^{\frac{D}{4}}}
C(g_0)(4n_0)^{\frac{D}{4}}.
\la{asgo2}
\ee
Comparing eqs.(\ref{asgo1}) and (\ref{asgo2}), one finds the normalization
constant
\be
C(g_0)=(4n_0)^{-\frac{D}{4}}.
\nonumber
\ee
Let us now consider the limit $u\to 0$. Taking into account the OPE
\bea
\s _{n_0N}(u,\bar u)\s _{g_0}[\k_1,\k_2](0)
&=&\frac{ C_{n_0N,g_0}^{g_{n_0N}g_0}(\k_1,\k_2)}
{|u|^{\frac{D}{8}+2\D _{g_0}[\k_1,\k_2]-2\D _{g_{n_0N}g_0}[\k_1+\k_2]}}
\s _{g_{n_0N}g_0}[\k_1+\k_2](0)\nonumber\\
&+&
\frac {C_{n_0N,g_0}^{g_0g_{n_0N}}(\k_1,\k_2)}
{|u|^{\frac{D}{8}+2\D _{g_0}[\k_1,\k_2]-2\D _{g_{n_0N}g_0}[\k_1+\k_2]}}
\s _{g_0g_{n_0N}}[\k_1+\k_2](0) +\cdots ,
\label{ope2}
\eea
one obtains
\bea
&&G_0(u,\bar u)\rightarrow \nonumber\\
&&\frac{ C_{n_0N,g_0}^{g_{n_0N}g_0}(\k_1,\k_2)}{|u|^{\frac{D}{8}+2\D _{g_0}
[\k_1,\k_2]
-2\D _{g_{n_0N}g_0}[\k_1+\k_2]}}
\langle\s_{ g_0^{-1}}[-\k_1,-\k_2](\infty )
\s_{n_0N}(1)\s_{g_{n_0N}g_0}[\k_1+\k_2](0)\rangle \nonumber\\
&&+
\frac{ C_{n_0N,g_0}^{g_0g_{n_0N}}(\k_1,\k_2)}{|u|^{\frac{D}{8}+2\D _{g_0}
[\k_1,\k_2]
-2\D _{g_{n_0N}g_0}[\k_1+\k_2]}}
\langle\s_{ g_0^{-1}}[-\k_1,-\k_2](\infty )
\s_{n_0N}(1)\s_{g_0g_{n_0N}}[\k_1+\k_2](0)\rangle .
\la{asgo3}
\eea
It is not difficult to show that the correlators
$\langle\s_{ g_0^{-1}}\s_{n_0N}\s_{g_{n_0N}g_0}\rangle$ and
$\langle\s_{ g_0^{-1}}\s_{n_0N}\s_{g_0g_{n_0N}}\rangle$ are equal to
$ C_{n_0N,g_0}^{g_0g_{n_0N}}$
and $ C_{n_0N,g_0}^{g_{n_0N}g_0}$ respectively, and, moreover, are equal to
each other. It follows from eqs.(\ref{invcor}) and (\ref{invcor1}), and from the
obvious symmetry property of the structure constant
$C_{n_0N,g_0}^{g_{n_0N}g_0}(-\k_1,-\k_2)=
C_{n_0N,g_0}^{g_{n_0N}g_0}(\k_1,\k_2)$:
\bea
\langle\s_{ g_0^{-1}}\s_{n_0N}\s_{g_{n_0N}g_0}\rangle =
\langle\s_{g_{n_0N}g_0}\s_{n_0N}\s_{ g_0^{-1}}\rangle
=\left\{ \begin{array}{c}
\langle\s_{g_{n_0N}g_0^{-1}}\s_{n_0N}\s_{ g_0}\rangle
=R^DC_{n_0N,g_0}^{g_0g_{n_0N}}\\
\langle\s_{g_0^{-1}g_{n_0N}}\s_{n_0N}\s_{ g_0}\rangle
=R^DC_{n_0N,g_0}^{g_{n_0N}g_0}\end{array}\right. .
\label{2}
\eea
Thus, the correlator $G_0(u,\bar u)$ in the limit $u\to 0$ is expressed through
the structure constant
$$
C(n_0,\k_1;N-n_0,\k_2)\equiv C_{n_0N,g_0}^{g_{n_0N}g_0}(\k_1,\k_2)
$$
as follows
\be
G_0(u,\bar u)\rightarrow
\frac{ 2R^DC^2(n_0,\k_1;N-n_0,\k_2)}{|u|^{\frac{D}{8}+2\D _{g_0}
[\k_1,\k_2]-2\D _{g_{n_0N}g_0}[\k_1+\k_2]}}.
\la{asgo4}
\ee
On the other hand, taking into account that in the limit $u\to 0$ the root
$x(u)$ behaves as
\bea
|x+\frac{n_0}{N-n_0}|\to
Nn_0^{\frac{N-2n_0}{N}}(N-n_0)^{\frac{2n_0-2N}{N}}|u|^{\frac{1}{N}},
\nonumber
\eea
one gets from eq.(\ref{soldif1})
\be
G_0(u,\bar u)\rightarrow
\frac{ 2^{\frac{D}{12}}R^DC(g_0)}{|u|^{\frac{D}{8}-\frac{2}{N}d_2}}
N^{-\frac{D}{12}+4d_2}(N-n_0)^{-\frac{D}{12}+4\frac{n_0-N}{N}d_2}
n_0^{\frac{D}{6}-4\frac{n_0}{N}d_2}.
\la{asgo5}
\ee
Comparing eqs.(\ref{asgo4}) and (\ref{asgo5}), one obtains the following
expression for the structure constant
\be
C(n_0,\k_1;N-n_0,\k_2)=2^{-\frac{5D+12}{24}}
N^{-\frac{D}{24}+2d_2}(N-n_0)^{-\frac{D}{24}-2\frac{N-n_0}{N}d_2}
n_0^{-\frac{D}{24}-2\frac{n_0}{N}d_2},
\la{strcons}
\ee
where
\be
d_2\equiv d_2(n_0,\k_1;N-n_0,\k_2)=\frac {D}{24}-
\frac {n_0}{8(N-n_0 )}(\k_2^2-\frac {D}{3})-
\frac {N-n_0}{8n_0}(\k_1^2-\frac {D}{3})+\frac {1}{4}\k_1\k_2 .
\nonumber
\ee
It is now not difficult to express any three-point correlator of the form
$\langle\s_{ g^{-1}g_{IJ}}\s_{IJ}\s_{g}\rangle$ through the structure constant
$C(n,\k ;m,\q )$. First of all let us note that any twist field 
$\s_g[\{\k_\alpha\} ]$ has the following decomposition into the tensor product 
of the twist fields 
$\s_{(n)}[\k ]$
\be
\s_g[\{\k_\alpha\} ]=\bigotimes_{\a =1}^{N_{str}} \, \s_{(n_\a )}[\k_\a ],
\la{1a}
\ee
where the element $g$ has the decomposition $(n_1)(n_2)\cdots (n_{N_{str}})$.
\footnote{ we will use the notation $(-n_1)(-n_2)\cdots (-n_{N_{str}})$ for the
decomposition of the element $g^{-1}$.}

\noindent Then, due to eq.(\ref{2}), the structure constant $C(n,\k ;m,\q )$
with arbitrary $n$ and $m$ is equal to
\be
C(n,\k ;m,\q )=R^{-D}\langle\s_{(-n-m)}[-\k -\q ](\infty )
\s_{IJ}(1)\s_{(n)}[\k ]\otimes\s_{(m)}[\q ](0)\rangle ,
\la{2a}
\ee
where $I\in (n)$ and $J\in (m)$.

\noindent
By using eqs.(\ref{1a}) and (\ref{2a}), one can easily get the following
expression for the three-point correlator
\bea
&&\langle\s_{ g^{-1}g_{IJ}}[\{\q_\alpha\} ](\infty )
\s_{IJ}(1)\s_{g}[\{\k_\alpha\} ](0)\rangle =
\langle\s_{g}[\{\k_\alpha\} ](\infty )
\s_{IJ}(1)\s_{ g^{-1}g_{IJ}}[\{\q_\alpha\} ](0)\rangle \nonumber\\
&&=\langle\s_{(-n_1-n_2)}[\q ]\bigotimes_{\a =3}^{N_{str}}\s_{(-n_\a )}[\q_\a ] 
(\infty )\s_{IJ}(1)
\s_{(n_1)}[\k_1 ]\otimes\s_{(n_2)}[\k_2 ]
\bigotimes_{\a =3}^{N_{str}}\s_{(n_\a )}[\k_\a ](0)\rangle\nonumber\\
&&=\prod_{\a =3}^{N_{str}}\d_R^D(\q_\a +k_\a )
\langle\s_{(-n_1-n_2)}[\q ](\infty )\s_{IJ}(1)
\s_{(n_1)}[\k_1 ]\otimes\s_{(n_2)}[\k_2 ](0)\rangle\nonumber\\
&&=C(n_1,\k_1;n_2,\k_2)\d_R^D(\q +k_1+\k_2 )
\prod_{\a =3}^{N_{str}}\d_R^D(\q_\a +k_\a ),
\la{3a}
\eea
where $I\in (n_1)$ and $J\in (n_2)$.

\noindent
It is now clear that the structure constant $C_{IJ,g}^{g_{IJ}g}$ in the OPE
of $\s_{IJ}$ and $\s_g$ is just equal to $C(n_1,\k_1;n_2,\k_2)$, and that the
structure constant $C_{IJ,g^{-1}g_{IJ}}^{g^{-1}}$ (which coincides with
$C_{IJ,g^{-1}g_{IJ}}^{g_{IJ}g^{-1}g_{IJ}}$ due to eq.(\ref{2})) in the OPE
\bea
&&\s _{IJ}(u,\bar u)\s _{g^{-1}g_{IJ}}[\{\q_\a \} ](0)
=\sum_{\q_1,\q_2}\frac{\d_{\q_1+\q_2-\q ,0}}
{|u|^{\frac{D}{8}+2\D _{g^{-1}g_{IJ}}[\{\q_\a \} ]-2\D _{g}[\{\q_\a \} ]}}
\nonumber\\
&&\times\left( C_{IJ,g^{-1}g_{IJ}}^{g^{-1}}(\q_1,\q_2)
\s _{g^{-1}}[\{\q_\a \} ](0)+C_{IJ,g^{-1}g_{IJ}}^{g^{-1}}(\q_1,\q_2)
\s _{g_{IJ}g^{-1}g_{IJ}}[\{\q_\a \} ](0)\right) +\cdots
\label{ope6}
\eea
is equal to
\bea
C_{IJ,g^{-1}g_{IJ}}^{g^{-1}}(\q_1,\q_2)=R^{-D}C(n_1,\q_1;n_2,\q_2).
\nonumber\eea
In particular, the structure constants 
$C_{n_\infty N,g_0}^{g_{n_\infty N}g_0}$ and 
$C_{n_0+n_\infty ,N;g_0}^{g_{n_0+n_\infty ,N}g_0}$, which will be used to find 
the normalization constant $C(g_0,g_\infty )$ are given by
\bea
\la{5a}
&&C_{n_\infty N,g_0}^{g_{n_\infty N}g_0}(\k_1,\k_2)=
R^{-D}C(n_\infty -n_0,\k_1;N-n_\infty ,\k_2),\\
&&C_{n_0+n_\infty ,N;g_0}^{g_{n_0+n_\infty ,N}g_0}(\k_1,\k_2)=
R^{-D}C(N-n_\infty -n_0,\k_1;n_\infty ,\k_2).\nonumber
\eea

Now we are ready to determine the normalization constant $C(g_0,g_\infty )$
by factorizing $G(u,\bar u)$ in the limit $u\to 0$ on tree-point
functions. According to eq.(\ref{ux}), $u\to 0$ in the following
three cases  
\bea 
\nonumber 
I)~~x\to -\frac{n_0}{N-n_0};~~~~ 
II)~~x\to \infty;~~~~
III)~~x\to \frac{N-n_0-n_{\infty}}{N-n_0}
\eea
and, conversely, any root $x_M=x_M(u)$ of eq.(\ref{ux}) tends to one of 
these values when $u\to 0$. Evidently, these three possible 
asymptotics correspond to three different choices of the indices $K$ 
and $L$ in eq.(\ref{s2}). 

Let us begin with the case $K=n_0$, $L=N$. By using the OPE
(\ref{ope2}) and the normalization (\ref{norm}) of two-point correlators, one
gets in the limit $u\to 0$
\be
G(u,\bar u)\rightarrow \d_R^D(\k_1+\k_2+\k_3+\k_4)
\frac{ C(n_0,\k_1;N-n_0,\k_2)C(n_\infty ,\k_3;N-n_\infty ,\k_4)}
{|u|^{\frac{D}{8}+2\D _{g_0}[\k_1,\k_2]-2\D _{g_{n_0N}g_0}[\k_1+\k_2]}}.
\la{7a}
\ee
In this case the root $x(u)$ has the following behaviour in the vicinity of
$u=0$
\be
|x+\frac{n_0}{N-n_0}|\to
Nn_0^{\frac{N-n_0}{N}}n_\infty^{-\frac{n_\infty}{N}}
(N-n_0)^{\frac{n_0-2N}{N}}(N-n_\infty )^{\frac{n_\infty }{N}}|u|^{\frac{1}{N}}.
\la{8a}
\ee
By using eqs.(\ref{soldif}) and (\ref{8a}), one can easily find 
\bea
&&G(u,\bar u)\rightarrow \d_R^D(\k_1+\k_2+\k_3+\k_4)
\frac{ C(g_0,g_\infty )}
{|u|^{\frac{D}{8}-\frac{2}{N}d_2}}
\left( \frac{n_0N(N-n_\infty )}{(N-n_0)^2(n_0-n_\infty )}\right)^{-\frac{D}{12}}
\times \nonumber\\
&&\times 
\left( \frac{n_0}{N-n_0}\right)^{2d_0}
\left( \frac{N}{N-n_0}\right)^{2d_1}
\left( \frac{n_0(N-n_\infty )}{(N-n_0)(n_0-n_\infty )}\right)^{2d_4}\nonumber\\
&&\times \left( \frac{N-n_\infty }{N-n_0}\right)^{2d_3}
\left( Nn_0^{\frac{N-n_0}{N}}n_\infty^{-\frac{n_\infty}{N}}
(N-n_0)^{\frac{n_0-2N}{N}}(N-n_\infty )^{\frac{n_\infty }{N}}\right)^{2d_2}.
\la{9a}
\eea
It is not difficult to verify that 
$$
\D _{g_0}[\k_1,\k_2]-\D _{g_{n_0N}g_0}[\k_1+\k_2]=-\frac{1}{N}d_2 ,
$$
as one should expect.

\noindent Comparing eqs.(\ref{7a}) and (\ref{9a}), one can obtain the
normalization constant. However, for general values of $D$ the corresponding
expression looks rather complicated and will not be written down. For $D=24$
one should take into account that the coefficients $d_i$ are given by the
following simple formulas
\bea
&&d_0=1+\frac 14 k_1k_4, \quad d_1=1+\frac 14 k_3k_4,
\quad d_2=1+\frac 14 k_1k_2, \nonumber\\
&&d_3=1+\frac 14 k_2k_3,\quad d_4=1+\frac 14 k_1k_3 ,
\la{11a}
\eea
where $k_ik_j\equiv \k_i\k_j-\frac 12 k_i^+k_j^--\frac 12 k_i^-k_j^+$.

\noindent By using eq.(\ref{11a}), one easily obtains
\be
C(g_0,g_\infty )=\frac{2^{-11}}{n_0(N-n_0)n_\infty (N-n_\infty )
(n_\infty -n_0)^2}
\left(\frac{ N-n_0}{ n_\infty -n_0}\right)^{2+\frac 12 (k_1+k_3)k_4}.
\la{12a}
\ee
Thus, we have found the normalization constant for $N$ correlators which are
presented in the first and second terms of eq.(\ref{s2}). 

Let us now determine the normalization constant for $n_\infty -n_0$ correlators
of the form  $\langle\s_{g_\infty (J)}
\s_{n_0J}\s_{n_\infty N}\s_{g_0}\rangle $. By using the OPE (\ref{ope6}) and
eq.(\ref{5a}), one finds in the limit $u\to 0$
\be
G(u,\bar u)\rightarrow \d_R^D(\k_1+\k_2+\k_3+\k_4)
\frac{ C(n_\infty -n_0,\k_2+\k_4;N-n_\infty ,-\k_4)
C(n_\infty -n_0,\k_2+\k_4;n_0 ,\k_1)}
{|u|^{\frac{D}{8}+2\D _{g_0}[\k_1,\k_2]-
2\D _{g_{n_\infty N}g_0}[\k_1,\k_2+\k_4,\k_4]}}.
\la{1c}
\ee
Taking into account the behaviour of the root $x(u)$ in the vicinity of $u=0$
\bea
|x|\to \left( (n_\infty -n_0)^{n_0-n_\infty}\frac{n_{\infty}^{n_{\infty}}}
{n_0^{n_0}}\left(\frac{N-n_0}{N-n_{\infty}}\right)^{N-n_{\infty}}\right)
^{\frac{1}{n_\infty -n_0}}|u|^{\frac{1}{n_0-n_\infty }},
\nonumber
\eea
one obtains from eq.(\ref{soldif})
\bea
&&G(u,\bar u)\rightarrow \frac{\d_R^D(\k_1+\k_2+\k_3+\k_4)C(g_0,g_\infty )}
{|u|^{\frac{D}{8}+\frac{2(d_0+d_1+d_2+d_3+d_4-\frac{D}{12})}{n_\infty -n_0}}}
\nonumber\\
&&\times
\left( (n_\infty -n_0)^{n_0-n_\infty}\frac{n_{\infty}^{n_{\infty}}}
{n_0^{n_0}}\left(\frac{N-n_0}{N-n_{\infty}}\right)^{N-n_{\infty}}\right)
^{\frac{2(d_0+d_1+d_2+d_3+d_4-\frac{D}{12})}{n_\infty -n_0}}.
\la{3c}
\eea
A simple calculation shows that
\bea
\D _{g_0}[\k_1,\k_2]-
\D _{g_{n_\infty N}g_0}[\k_1,\k_2+\k_4,\k_4]=
\frac{d_0+d_1+d_2+d_3+d_4-\frac{D}{12}}{n_\infty -n_0}.
\nonumber
\eea
The normalization constant $C(g_0,g_\infty )$ can be now found from
eqs.(\ref{1c}) and (\ref{3c}). For $D=24$ the computation drastically
simplifies if one notes that
\bea
&&d_0+d_1+d_2+d_3+d_4-2=-1-\frac 14 k_1k_3,\nonumber\\
&&d_2(n_\infty -n_0,\k_2+\k_4;N-n_\infty ,-\k_4)=-\frac{N-n_0}{n_\infty -n_0}
(1+\frac 14 k_1k_3),\nonumber\\
&&d_2(n_\infty -n_0,\k_2+\k_4;n_0 ,\k_1)=
-\frac{n_\infty}{n_\infty -n_0}
(1+\frac 14 k_1k_3).\nonumber
\eea
Then, one can easily show that $C(g_0,g_\infty )$ is again given by
eq.(\ref{12a}).

The normalization constant for the remaining $N-n_0-n_\infty$ correlators of
the form 
$$\langle\s_{g_\infty (J)}\s_{n_0J}\s_{n_0+n_\infty ,N}\s_{g_0}\rangle $$
can be found in the same manner and is again defined by eq.(\ref{12a}).

Up to now we considered the correlators
$G_{IJKL}(u,\bar u)=
\langle\s_{g_\infty }(\infty )\s_{IJ}(1)\s_{KL}(u,\bar u)\s_{g_0}(0)\rangle $ 
with $|u|<1$. The correlators $G_{IJKL}(u,\bar u)$ with
$|u|>1$ can be calculated in the same way, and their dependence on $u$ is given
by eq.(\ref{soldif}) as well. The normalization constant in this case is
derived 
by studying the limit $u\to\infty$ and coincides with the previously found
constant (\ref{12a}). The time-ordering, therefore, can be omitted, and to
complete the computation of the S-matrix element we have to integrate the
correlator $F(u,\bar u)$ (\ref{fuu}) over the complex plane. With the help of the momentum
conservation law, the mass-shell condition and eq.(\ref{11a}), one can rewrite
eq.(\ref{soldif})  in the following form
\bea
G_{IJKL}(u,\bar u)&=&\d_R^D(\k_1+\k_2+\k_3+\k_4)C(g_0,g_\infty )(n_\infty -n_0)^2
|\frac{du}{dx}|^{-2}|u|^{-1}
\frac {|x-\alpha_1|^2|x-\alpha_2|^2}
{|x-\frac {n_0}{n_0-n_\infty }|^4}\nonumber\\
&\times&
\left|\frac{x(x-\frac {N-n_0-n_\infty }{N-n_0})}{x-\frac {n_0}{n_0-n_\infty }}
\right|^{\frac 12 k_1k_4}
\left|\frac 
{(x-1)(x+\frac {n_0}{N-n_0})}{x-\frac {n_0}{n_0-n_\infty }}
\right|^{\frac 12 k_3k_4},\nonumber
\eea 
where the equality
\bea
\frac {1}{u}\frac {du}{dx}=
\frac {(n_0-n_\infty )(x-\a_1)^2(x-\a_2)^2}
{x(x-1)(x-\frac {N-n_0-n_\infty }{N-n_0})
(x-\frac {n_0}{n_0-n_\infty })(x+\frac {n_0}{N-n_0})}\nonumber
\eea
was used.

\noindent Now the integral $\int d^2u|u|G_{IJKL}(u,\bar u)$ can be easily 
calculated by changing the variables $u\to x$
\bea
&&\int d^2u|u|G_{IJKL}(u,\bar u)=
\d_R^D(\k_1+\k_2+\k_3+\k_4)C(g_0,g_\infty )(n_\infty -n_0)^2\nonumber\\
&&\int_{V_{IJKL}} d^2x
\frac {|x-\alpha_1|^2|x-\alpha_2|^2}
{|x-\frac {n_0}{n_0-n_\infty }|^4}
\left|\frac{x(x-\frac {N-n_0-n_\infty }{N-n_0})}{x-\frac {n_0}{n_0-n_\infty }}
\right|^{\frac 12 k_1k_4}
\left|\frac 
{(x-1)(x+\frac {n_0}{N-n_0})}{x-\frac {n_0}{n_0-n_\infty }}
\right|^{\frac 12 k_3k_4},
\la{55}
\eea 
where we have taken into account that under this change of variables the
$u$-sphere is mapped onto the domain $V_{IJKL}$.

\noindent Since the correlator $F(u,\bar u)$ is equal to the sum
$$
F(u,\bar u)=\frac {C_0C_\infty }{N!}2n_0(N-n_0)n_\infty (N-n_\infty )
\sum_{IJKL}G_{IJKL}(u,\bar u),$$
where the summation goes over the set of indices listed in eq.(\ref{s2}),
the integral $\int d^2u|u|F(u,\bar u)$ is equal to
\bea
&&\int d^2u|u|F(u,\bar u)=\frac{2^{-10}\d_R^D(\k_1+\k_2+\k_3+\k_4)}
{\sqrt {n_0(N-n_0)n_\infty (N-n_\infty )}}
\left(\frac{ N-n_0}{ n_\infty -n_0}\right)^{2+\frac 12 (k_1+k_3)k_4}
\nonumber\\
&&\int d^2x
\frac {|x-\alpha_1|^2|x-\alpha_2|^2}
{|x-\frac {n_0}{n_0-n_\infty }|^4}
\left|\frac{x(x-\frac {N-n_0-n_\infty }{N-n_0})}{x-\frac {n_0}{n_0-n_\infty }}
\right|^{\frac 12 k_1k_4}
\left|\frac 
{(x-1)(x+\frac {n_0}{N-n_0})}{x-\frac {n_0}{n_0-n_\infty }}
\right|^{\frac 12 k_3k_4},
\la{555}
\eea 
Finally, performing the change of variables
$$
\frac {n_\infty -n_0}{N-n_0} z=\frac 
{x(x-\frac {N-n_0-n_\infty }{N-n_0})}{x-\frac {n_0}{n_0-n_\infty }},$$
one gets the following expression for the integral
\be
\int d^2u|u|F(u,\bar u)=
\frac{2^{-9}\d_R^D(\k_1+\k_2+\k_3+\k_4)}{\sqrt {n_0(N-n_0)n_\infty 
(N-n_\infty )}}\int d^2z
|z|^{\frac 12 k_1k_4}|1-z|^{\frac 12 k_3k_4}.
\la{66}
\ee 
The S-matrix element can be now found by using eq.(\ref{matel3})  
and by taking the limit $R\to\infty$:
\bea
\langle f|S|i\rangle &=&-i\frac{\lambda^22^{-8}
N^3\d (k_1^-+k_2^-+k_3^-+k_4^-)\d^D(\k_1+\k_2+\k_3+\k_4)}
{\sqrt {n_0(N-n_0)n_\infty (N-n_\infty )}}
\int d^2z|z|^{\frac 12 k_1k_4}|1-z|^{\frac 12 k_3k_4}\nonumber\\
&=&-i\frac{\lambda^22^{-8}
N\d (\sum_i k_i^-)\d^D(\sum_i \k_i)}
{\sqrt {k_1^+k_2^+k_3^+k_4^+}}
\int d^2z|z|^{\frac 12 k_1k_4}|1-z|^{\frac 12 k_3k_4}.
\label{77}
\eea
Let us now represent the light-cone momenta $k_i^+$ as $k_i^+=\frac {m_i}{N}$
and rewrite eq.(\ref{77}) in the following form
\bea
\langle f|S|i\rangle &=&-i\frac{\lambda^22^{-8}
N\d_{m_1+m_2+m_3+m_4,0}\d (\sum_i k_i^-)\d^D(\sum_i \k_i)}
{\sqrt {k_1^+k_2^+k_3^+k_4^+}}
\int d^2z|z|^{\frac 12 k_1k_4}|1-z|^{\frac 12 k_3k_4}.
\label{88}
\eea
In the limit $N\to\infty$ the combination $N\d_{m_1+m_2+m_3+m_4,0}$ goes to 
$\d (\sum_i k_i^+)$ and eq.(\ref{88}) acquires the form
\bea
\langle f|S|i\rangle &=&-i\frac{\lambda^22^{-9}
\d^{D+2}(\sum_i k_i^\mu )}
{\sqrt {k_1^+k_2^+k_3^+k_4^+}}
\int d^2z|z|^{\frac 12 k_1k_4}|1-z|^{\frac 12 k_3k_4}.
\label{99}
\eea
Taking into account that the scattering amplitude $A$ is related to the
S-matrix as follows (see e.g. \cite{GSW})
$$
\langle f|S|i\rangle =-i\frac{\d^{D+2}(\sum_i k_i^\mu )}
{\sqrt {k_1^+k_2^+k_3^+k_4^+}}A(1,2,3,4),$$
one finally gets
$$
A(1,2,3,4)=\lambda^22^{-9}\int d^2z|z|^{\frac 12 k_1k_4}|1-z|^{\frac 12
k_3k_4}$$
that is the well-known Virasoro amplitude.

\section{Conclusion}
In this paper we developed the technique for calculating scattering amplitudes
of bosonic string states by using the interacting $S^N\R^{24}$ orbifold sigma
model. The scattering amplitude turned out to be automatically
Lorentz-invariant. It gives a strong evidence that the corresponding
two-dimensional Yang-Mills model should possess the same invariance.

It would be of interest to trace the appearance
of the loop amplitudes in the framework of the $S^N\R^{24}$ orbifold 
sigma model. Obviously, the one-loop amplitude requires the 
computation of the correlator of four $Z_2$-twist fields sandwiched
between the asymptotic states that technically results in constructing
the non-commutative Green functions in the presence of six twist
fields. We note that cancellation of possible divergences in the 
amplitude may require the further perturbation of the CFT action by 
higher-order contact terms.

The next important problem to be solved is to consider the $S^N\R^{8}$
supersymmetric orbifold sigma model and to prove the DVV conjecture. It is not
difficult to introduce twist fields for fermionic variables and to calculate
their conformal dimensions. However, the calculation of four-point correlators
of the twist fields is more complicated problem and is now under consideration.
It is not excluded that the simplest way to solve the problem is to bosonize
the fermion fields.

{\bf ACKNOWLEDGMENT} The authors thank I.Y.Aref'eva, L.O.Chekhov, P.B.Medvedev
and N.A.Slavnov for valuable discussions. One of the authors (S.F.) is
grateful to Professor J.Wess for kind hospitality and the Alexander von 
Humboldt Foundation for the support. This work has been supported in part 
by the RFBI grants N96-01-00608, N96-01-00551 and by the ISF 
grant a96-1516.  

\setcounter{section}{0}
\appendix{}
\setcounter{equation}{0}
In this Appendix we consider some properties of the map (\ref{map}) and outline
the derivation of the differential equation (\ref{difur4}) for the four-point
correlators (\ref{guu}). 

Let us consider the map (\ref{map}) 
\be 
z=\frac {t^{n_0}(t-t_0)^{N-n_0}}{(t-t_\infty )^{N-n_\infty}} 
\frac {(t_1-t_\infty )^{N-n_\infty}}{t_1^{n_0}(t_1-t_0)^{N-n_0}} 
\equiv u(t).  
\la{a1} 
\ee 
This map is the $N$-fold covering of the $z$-sphere by the 
$t$-sphere. Obviously, it branches at the points $t=0,t_0,t_\infty$ 
and $\infty$. To find other branch 
points we have to solve the following equation:  \bea \frac{d\log 
z}{dt}&=& 
\frac{n_0}{t}+\frac{N-n_0}{t-t_0}-\frac{N-n_\infty}{t-t_\infty}\nonumber\\
&=&\frac{n_\infty t^2+\left( (N-n_0-n_\infty )t_0-Nt_\infty \right) t+
n_0t_0t_\infty}{t(t-t_0)(t-t_\infty )}.
\la{a2}
\eea
In general there are two different solutions $t_1$ and $t_2$ of this equation,
and the map (\ref{a1}) has the following form in the vicinity of these points
\bea
z-z_i\sim (t-t_i)^2 ,\quad z_1=1=u(t_1),\quad z_2=u=u(t_2).
\nonumber
\eea
Due to the projective transformations, we can impose three relations on
positions of branch points. However, we have already chosen the points $0$ and
$\infty$ as two branch points, therefore, only one relation remains to be
imposed. Since the differential equation on the four-point correlator is
written with respect to the point $u$, it is convinient not to fix the position
of the point $t_2\equiv x$. Then, the remaining relation that leads to the
rational dependence of points $t_0,t_\infty$ and $t_1$ on $x$ looks as follows
\be
t_0=x-1.
\la{a4}
\ee
The point $x$ is supposed to be a solution of eq.(\ref{a2}). Therefore, one can
immediately derive from eqs.(\ref{a2}) and (\ref{a4}) that  $t_\infty$ is 
expressed through the point $x$ as
\be
t_\infty =x-\frac {(N-n_\infty )x}{(N-n_0)x+n_0}.
\la{a5}
\ee
The second solution of eq.(\ref{a2}) can be now easily found and is given by
\bea
t_1&=&\frac {N-n_0-n_\infty }{n_\infty}+\frac {n_0x}{n_\infty}-
\frac {N(N-n_\infty )x}{n_\infty ((N-n_0)x+n_0)}\nonumber\\
&=&\frac{n_0(x-1)\left( (N-n_0)x+n_0+n_\infty -N\right)}
{n_\infty\left( (N-n_0)x+n_0\right)}.
\la{a6}
\eea
The rational function $u(x)$ is defined by the following equation
\be
u(x)=\frac {x^{n_0}(x-t_0)^{N-n_0}(t_1-t_\infty )^{N-n_\infty}}
{(x-t_\infty )^{N-n_\infty}t_1^{n_0}(t_1-t_0)^{N-n_0}}.
\la{a7}
\ee
By using eqs.(\ref{a4}),(\ref{a5}) and (\ref{a6}), one can derive the following
relations
\bea
t_1-t_0&=&\frac{(N-n_0)(x-1)\left( (n_0-n_\infty )x-n_0\right)}
{n_\infty\left( (N-n_0)x+n_0\right)},\nonumber\\
t_1-t_\infty &=&\frac{\left( (n_0-n_\infty )x-n_0\right)
\left( (N-n_0)x+n_0+n_\infty -N\right)}
{n_\infty\left( (N-n_0)x+n_0\right)}.\nonumber
\eea
Then the rational function $u(x)$ is found to be equal to
\bea
u=u(x)&=&(n_0-n_{\infty})^{n_0-n_{\infty}}\frac{n_{\infty}^{n_{\infty}}}
{n_0^{n_0}}\left(\frac{N-n_0}{N-n_{\infty}}\right)^{N-n_{\infty}}
\left(\frac{x+\frac{n_0}{N-n_0}}{x-1}\right)^N\nonumber\\
&\times&\left(\frac{x-\frac{N-n_0-n_{\infty}}{N-n_0}}{x}\right)^{N-n_0-n_{\infty}}
\left(x-\frac{n_0}{n_0-n_{\infty}}\right)^{n_0-n_{\infty}}.
\la{a10}
\eea
To obtain the differential equation (\ref{difur4}) we need to know the
decomposition of the roots $t_K(z)$ and $t_L(z)$ in the vicinity of $z=u$. Let
us take the logarithm of the both sides of eq.(\ref{a1}):
\be
\log {\frac zu}=n_0\log {\frac tx}+(N-n_0)\log {\frac {t-t_0}{x-t_0}}-
(N-n_\infty )\log {\frac {t-t_\infty }{x-t_\infty }}.
\la{a11}
\ee
Decomposition of the l.h.s. of eq.(\ref{a11}) around $z=u$ and the r.h.s. 
of eq.(\ref{a11}) around $t=x$ gives:
\be
\sum_{k=1}^\infty \frac{(-1)^{k+1}}{k}\left(\frac{z-u}{u}\right)^k=
(t-x)^2\sum_{k=0}^\infty a_k(t-x)^k ,
\la{a12}
\ee
where the coefficients $a_k$ are equal to
\be
a_k=\frac {(-1)^{k-1}}{k+2}\left( \frac {n_0}{x^{k+2}}+
\frac {N-n_0}{(x-t_0)^{k+2}}-\frac {N-n_\infty}{(x-t_\infty)^{k+2}}\right) .
\la{a13}
\ee
It is clear from eq.(\ref{a12}) that $t(z)$ has the following decomposition
\be
t-x=\sum_{k=1}^\infty c_k(z-u)^{\frac k2}.
\la{a14}
\ee
Substituting eq.(\ref{a14}) into eq.(\ref{a12}), one finds
\bea
&&c_1^2=\frac {1}{ua_0},\quad c_2=-\frac {a_1}{2ua_0},\nonumber\\
&&2a_0c_1c_3=-\frac {1}{2u^2}+\frac {5a_1^2}{4u^2a_0^3}-
\frac {a_2}{u^2a_0^2}.
\la{a15}
\eea
Next coefficients are not important for us.

\noindent Then, by using the decomposition (\ref{a14}) and eq.(\ref{a15}), one
gets
\bea
&&\left(\frac {t''}{t'}\right) '=\frac {1}{2(z-u)^2}+O(1),\nonumber\\
&&\left(\frac {t''}{t'}\right)^2=\frac {1}{4(z-u)^2}+\frac {3}{z-u}
\left(\frac {c_2^2}{c_1^2}-\frac {c_3}{c_1}\right) +O(1),\nonumber\\
&&\frac {c_2^2}{c_1^2}-\frac {c_3}{c_1}=
\frac {1}{4u}\left( 1+\frac {2a_2}{a_0^{2}}-\frac {3a_1^2}{2a_0^3}\right) .
\nonumber
\eea
Finally, taking into account that in the set of $N$ roots $t_M(z)$ only two
roots $t_K(z)$ and $t_L(z)$ have the decomposition (\ref{a14}), we obtain
eqs.(\ref{dect}) and (\ref{difur3}).

The coefficients $a_k$ can be rewritten as the following functions of $x$:
\bea
a_0&=&\frac {n_0(n_0+n_\infty -N)}{2(N-n_\infty )x^2}+
\frac {n_0(N-n_0)}{(N-n_\infty )x}+
\frac {(N-n_0)(n_\infty -n_0)}{2(N-n_\infty )}\nonumber\\
&=&\frac {(N-n_0)(n_\infty -n_0)}{2(N-n_\infty )x^2}(x-\a_1)(x-\a_2),
\la{a18}\\
a_1&=&\frac {n_0((N-n_\infty )^2-n_0^2)}{3(N-n_\infty )^2x^3}-
\frac {n_0^2(N-n_0)}{(N-n_\infty )^2x^2}\nonumber\\&-&
\frac {n_0(N-n_0)^2}{(N-n_\infty )^2x}+
\frac {(N-n_0)((N-n_\infty )^2-(N-n_0)^2)}{3(N-n_\infty )^2},\nonumber\\
a_2&=&-\frac {n_0((N-n_\infty )^3-n_0^3)}{4(N-n_\infty )^3x^4}+
\frac {n_0^3(N-n_0)}{(N-n_\infty )^3x^3}+
\frac {3n_0^2(N-n_0)^2}{2(N-n_\infty )^3x^2}\nonumber\\
&+&\frac {n_0(N-n_0)^3}{(N-n_\infty )^3x}-
\frac {(N-n_0)((N-n_\infty )^3-(N-n_0)^3)}{4(N-n_\infty )^3}.\nonumber
\eea
To obtain the differential equation (\ref{difur4}) we have to use the following
important equalities on $\frac {1}{u}\frac {du}{dx}$, that can be derived by
using eqs.(\ref{a10}) and (\ref{a18})
\bea
\frac {1}{u}\frac {du}{dx}&=&\frac {n_0+n_\infty -N}{x}
-\frac {N}{x-1}+\frac {N}{x+\frac {n_0}{N-n_0}}\nonumber\\
&+&
\frac {N-n_0-n_\infty }{x-\frac {N-n_0-n_\infty }{N-n_0}}+
\frac {n_0-n_\infty }{x-\frac {n_0}{n_0-n_\infty }},\nonumber\\
\frac {1}{u}\frac {du}{dx}&=&\frac {4(N-n_\infty )^2x^4a_0^2}
{(N-n_0)^2(n_0-n_\infty )x(x-1)(x-\frac {N-n_0-n_\infty }{N-n_0})
(x-\frac {n_0}{n_0-n_\infty })(x+\frac {n_0}{N-n_0})}\nonumber\\
&=&\frac {(n_0-n_\infty )(x-\a_1)^2(x-\a_2)^2}
{x(x-1)(x-\frac {N-n_0-n_\infty }{N-n_0})
(x-\frac {n_0}{n_0-n_\infty })(x+\frac {n_0}{N-n_0})}.\nonumber
\eea
Finally, to get eq.(\ref{difur4}) one should use the Lagrange 
interpolation formula for the ratio of two polynomials 
\bea 
\frac{P(x)}{Q(x)}=\sum_i\frac{P(x_i)}{Q'(x_i)}\frac{1}{x-x_i},
\nonumber
\eea
where $x_i$ are the simple roots of $Q(x)$ and $degP<degQ$.

These equalities drastically simplify  the derivation of eq.(\ref{difur4}).



\begin{thebibliography}{99}
{\small
\bibitem{BFSS} T.Banks, W.Fischler, S.H.Shenker, and L.Susskind, 
``M Theory as a Matrix Model: A Conjecture,''
hep-th/9610043.
\bibitem{T} W.Taylor, ``D-brane Field Theory on Compact Spaces,''
hep-th/9611042. 
\bibitem{M} L.Motl, ``Proposals on Nonperturbative Superstring Interactions,''
hep-th/9701025.
\bibitem{BS}T.Banks and N.Seiberg, ``Strings from Matrices,'' hep-th/9702187.
\bibitem{DVV}  R.Dijkgraaf, E.Verlinde and H.Verlinde, ``Matrix String Theory,''
hep-th/9703030.
\bibitem{DMVV} R.Dijkgraaf, G.Moore, E.Verlinde and H.Verlinde,
``Elliptic Genera of Symmetric Products and Second Quantized Strings,''
hep-th/9608096.
\bibitem{R} S.-J.Rey, ``Heterotic M(atrix) strings and Their Interactions,'' 
hep-th/9704158. 
\bibitem{DHVW1} L.Dixon, J.A.Harvey, C.Vafa and E.Witten, Nucl.Phys. {\bf
B261} (1985) 678.
\bibitem{DHVW2} L.Dixon, J.A.Harvey, C.Vafa and E.Witten, Nucl.Phys. {\bf
B274} (1986) 285.
\bibitem{BPZ} A.A.Belavin, A.M.Polyakov and A.B.Zamolodchikov, Nucl.Phys. {\bf
B241} (1984) 333.
\bibitem{DFMS} L.Dixon, D.Friedan, E.Martinec and S.Shenker, Nucl.Phys. {\bf
B282} (1987) 13.
\bibitem{GSW}
M.B. Green, J.H. Schwarz, E. Witten, 
Superstring Theory (Cambridge University Press, 1987).} 
\end{thebibliography}
\end{document}